\documentclass{article}

\usepackage[preprint]{neurips_2026}

\usepackage[utf8]{inputenc}
\usepackage[T1]{fontenc}
\usepackage{amsmath}
\usepackage{amssymb}
\usepackage{hyperref}
\usepackage{amsthm}
\usepackage{booktabs}
\usepackage{graphicx}
\usepackage{makecell}
\usepackage{mathtools}
\usepackage{algorithm}
\usepackage{microtype}
\usepackage{multirow}
\usepackage{nicefrac}
\usepackage{subfigure}
\usepackage{url}
\usepackage{wrapfig}
\usepackage{algpseudocode}
\usepackage{enumitem}
\usepackage{xcolor}
\usepackage[capitalize,noabbrev]{cleveref}

\title{Accelerating Suffix-Based Jailbreak Evaluation with Prefix-Shared KV Cache}

\author{%
  Xinhai Wang$^{1,*}$ \quad
  Shaopeng Fu$^{1,*}$ \quad
  Shu Yang$^{1}$ \quad
  Liangyu Wang$^{1}$ \quad
  Tianhang Zheng$^{2}$ \quad
  Di Wang$^{1}$ \\
  $^{1}$King Abdullah University of Science and Technology (KAUST) \quad
  $^{2}$Zhejiang University \\
  \texttt{\{xinhai.wang, shaopeng.fu, shu.yang, liangyu.wang, di.wang\}@kaust.edu.sa} \\
  \texttt{zthzheng@zju.edu.cn}
  \thanks{$*$Equal contribution.}
}

\begin{document}

\maketitle

\begin{abstract}
Suffix jailbreak attacks are a systematic tool for red-teaming large language models (LLMs), but they require evaluating many candidate suffixes before finding an effective jailbreak. This paper presents Prefix-Shared KV Cache (PSKV), a plug-and-play inference optimization tailored to suffix jailbreak evaluation. PSKV exploits the structure that many candidate prompts share the same harmful instruction prefix while differing only in the candidate suffix. Instead of redundantly recomputing or physically duplicating the prefix cache, PSKV stores one compact prefix KV cache and expands it lazily at each layer during candidate evaluation. This design preserves PyTorch autograd compatibility and enables larger batched searches with lower memory overhead. Across six suffix attacks and five open-source LLMs, PSKV reduces inference time by about 40\% and peak memory by about 50\% while preserving attack effectiveness.
\end{abstract}

\section{Introduction}

Suffix jailbreak attacks, which aim to synthesize jailbreak suffixes to bypass safety guardrails and elicit harmful content, have become a vital component for red-teaming Large Language Models (LLMs)~\citep{zou2023universal,chao2024jailbreakingblackboxlarge}.
Current approaches primarily employ heuristic search~\citep{zou2023universal,zhu2023autodan} or LLM-based generators~\citep{paulus2024advprompter} to generate these suffixes.
However, a major bottleneck of these methods is their prohibitively high computational cost.
Standard optimization-based attacks must iteratively evaluate a vast number of \textit{candidate} suffixes against the target LLM to identify effective jailbreaks.
While LLM-based generators offer faster test-time suffix generations, they merely shift the computational burden to the training phase, which still necessitates time-consuming suffix generation to construct training data.

Ideally, Key-Value (KV) caching could accelerate this evaluation process by reusing intermediate tensors.
However, in jailbreak scenarios, the instruction prefix is typically far longer than the suffix.
Standard KV cache implementations must duplicate this long prompt's cache for every candidate suffix in a batch, inducing excessive memory consumption that limits batch size.
General-purpose serving systems such as vLLM~\citep{kwon2023efficient} and SGLang~\citep{zheng2024sglang} improve KV-cache utilization through paging, scheduling, and request-level prefix reuse, but they target online forward-pass serving rather than differentiable, synchronous attack optimization loops.
Their custom serving runtimes operate outside the standard PyTorch autograd graph, restricting their utility to a subset of gradient-free attacks (\textit{e.g.}, BEAST attack~\citep{sadasivan2024fast}).
Consequently, these limitations collectively impede the effective acceleration of general suffix jailbreak attacks.

To overcome these limitations, we propose \textbf{Prefix-Shared KV Cache (PSKV)}, a lightweight, workload-aware, and plug-and-play\footnote{A step-by-step guide for quickly integrating PSKV into attack code is provided in Appendix~\ref{app:using-pskv}.} framework tailored for accelerating suffix jailbreak attacks.
Our method is motivated by a key \textit{structural redundancy} in jailbreak optimization: while a batch of candidate suffixes varies, they typically share an identical targeted harmful instruction as the prefix.
Standard KV cache implementations ignore this property, redundantly duplicating the long prefix cache for every candidate.
In contrast, PSKV stores a single, shared KV cache for the invariant prefix and reuses it across candidate suffixes through layer-wise lazy expansion.
By employing a suffix-centric alignment strategy, PSKV supports dynamic expansion and parallel batched loss computation.
Unlike encapsulated inference engines, PSKV integrates seamlessly into existing attack pipelines, achieving substantial acceleration and lower memory while preserving gradient access.

We evaluated PSKV across six widely used suffix jailbreak attacks and five popular open-source models.
We compared our framework against a baseline without KV cache and a standard KV cache implementation.
The results demonstrate that PSKV reduces inference time by approximately $40\%$ compared to the no-cache baseline and decreases peak memory usage by $50\%$ compared to the standard KV cache implementation.
Furthermore, because PSKV is a drop-in optimization that does not alter the high-level logic of the attack, it maintains the original Attack Success Rate (ASR).
This efficiency enables researchers to scale evaluations to larger models and broader search spaces.

In summary, our contributions are:
\begin{itemize}[leftmargin=20pt, itemsep=1pt, topsep=0pt, parsep=0pt, partopsep=0pt]
    \item We identify a mismatch between standard KV-cache execution and the shared-prefix, candidate-expanded structure of suffix-based jailbreak evaluation.
    \item We propose PSKV, a memory-efficient framework that combines prefix reuse, layer-wise lazy expansion, and suffix-centric tensor alignment to eliminate redundant prefix computation while preserving autograd support.
    \item We conduct evaluations demonstrating that PSKV substantially reduces both computational and memory costs, facilitating scalable evaluation of suffix-based jailbreak attacks.
\end{itemize}
\vspace{-10pt}
\section{Related Work}

\textbf{Automated Suffix Jailbreaking.}
Recent research has transitioned from manual red-teaming to automated adversarial optimization.
The foundational method, GCG~\citep{zou2023universal}, employs a greedy, gradient-guided search to iteratively refine an adversarial suffix. Building on this, subsequent works have explored more sophisticated search strategies. BEAST~\citep{sadasivan2024fast} replaces the greedy algorithm with beam search to explore a wider range of promising candidates. Zhu's AutoDAN~\citep{zhu2023autodan} utilizes genetic algorithms to evolve suffixes, mitigating the local optima problem of purely greedy methods. Other variants, such as GCQ (Greedy Coordinate Query)~\citep{hayase2024query}, extend coordinate-search-style optimization to a query-based setting by maintaining a buffer of promising candidates and using proxy/true losses to guide candidate selection.

An alternative paradigm uses LLMs themselves as generators. AdvPrompter~\citep{paulus2024advprompter} fine-tunes a language model to automatically compose adversarial prompts. Similarly, AmpleGCG trains a generative model to directly output effective suffixes, amortizing the search cost into a one-time training process~\citep{liao2024amplegcg}. Although these methods differ, they all require evaluating a large number of candidate suffixes to create a training set during either the model training or data generation phases. Our work presents an orthogonal optimization that accelerates this critical evaluation step, supporting scalable evaluation of both categories.

\textbf{Efficient Inference \& KV Caching.}
Optimizing Key-Value (KV) cache management is central to high-throughput LLM serving.
vLLM~\citep{kwon2023efficient} introduced PagedAttention to mitigate memory fragmentation in multi-user scenarios, while SGLang~\citep{zheng2024sglang} employs RadixAttention to organize reusable prompt states in a radix tree and reuse overlapping request prefixes in online serving.
ChunkAttention~\citep{ye2024chunkattention} further optimizes memory locality.

General-purpose serving engines such as vLLM and SGLang target online LLM serving, where requests arrive asynchronously and the system manages scheduling, paging, eviction, and request-level batching. Adversarial suffix optimization has a different structure: it repeatedly scores many candidate suffixes under the same prefix in a synchronous inner loop, often requiring gradients with respect to suffix embeddings. Naively expanding prefix KV tensors across candidates causes memory to grow with both candidate count and prefix length, while serving-engine cache reuse is typically implemented outside the PyTorch autograd graph and is not directly suitable for gradient-based attacks. Moreover, applying serving engines to adversarial optimization can introduce additional overhead and system complexity, such as scheduler latency or tree-lookup costs, that are unnecessary for this fixed shared-prefix workload.

PSKV is designed for this attack-specific workload. It uses \emph{layer-wise lazy expansion} to keep one shared prefix cache and avoids persistent candidate-expanded prefix caches, and \emph{suffix-centric tensor alignment} to support vectorized candidate substitution, gradient extraction, and scoring across ragged multi-prompt batches. Because PSKV is implemented with standard PyTorch operations, it preserves autograd support while avoiding fully expanded prefix-cache storage.
\section{Preliminaries}
\label{sec:preliminaries}

\begin{wrapfigure}{r}{0.45\textwidth}
    \vspace{-10pt}
  \includegraphics[width=0.45\columnwidth]{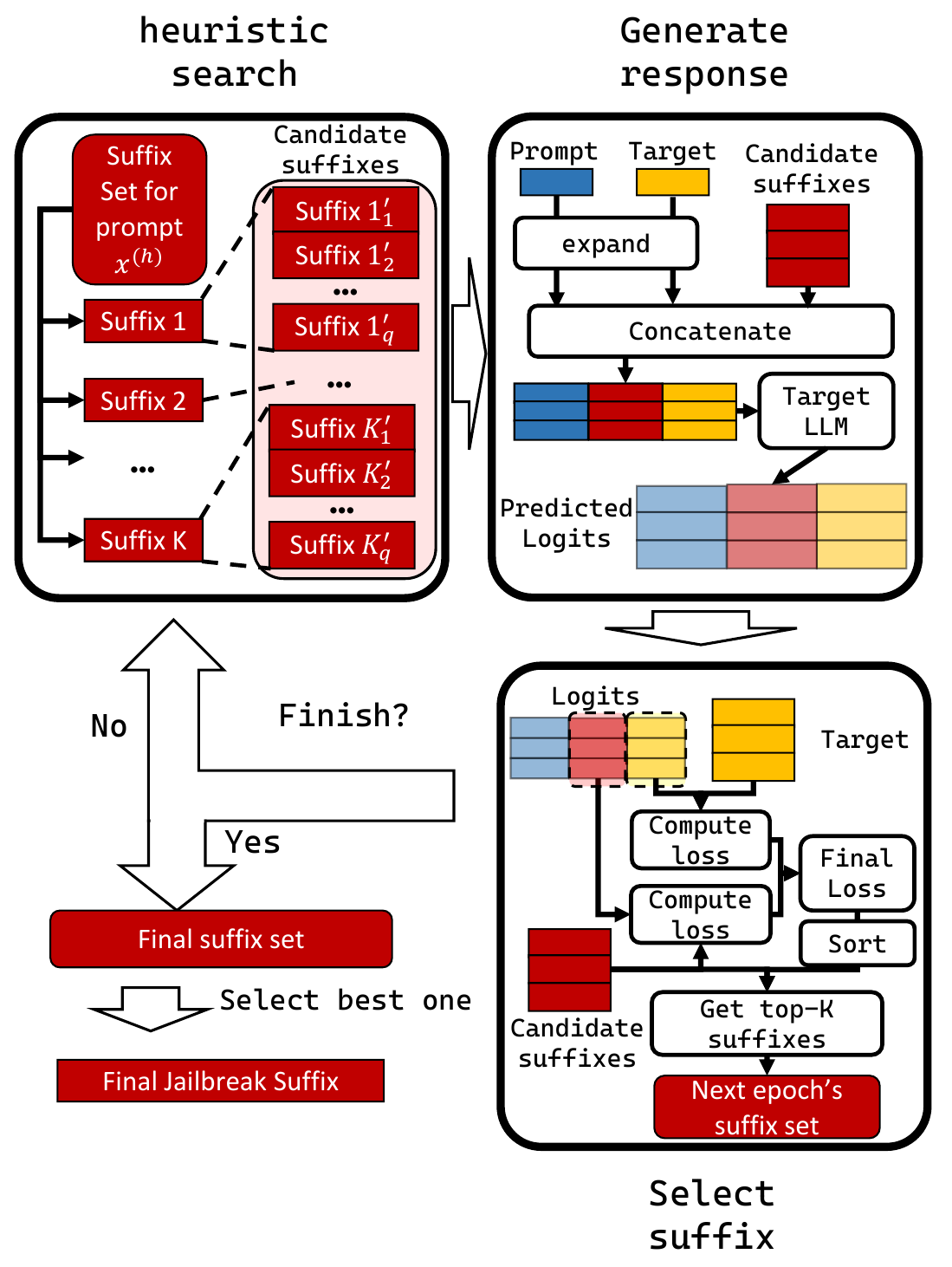}
  \vspace{-10pt}
  \caption{Computational pipeline of iterative suffix attacks, where many candidate suffixes are repeatedly scored under the same instruction prefix.}
  \label{fig:general_process}
  \vspace{-18pt}
\end{wrapfigure}

Fig.~\ref{fig:general_process} illustrates the iterative candidate-generation and scoring pipeline used by suffix-based jailbreak attacks. This pipeline forms the operational basis for the optimization problem defined below.

\textbf{Adversarial Suffix Optimization.}
We consider an LLM as a conditional distribution $p_{\theta}(y|x)$ over a vocabulary $\mathcal{V}$.
Given a harmful query $x^{(h)}$ and a target response $y^{(h)}$, the goal of a suffix attack is to find an adversarial sequence $x^{(s)} \in \mathcal{V}^L$ that maximizes the target probability:
\begin{align}
    x^{(s)*} = \mathop{\arg\max}_{x^{(s)} \in \mathcal{V}^L} \log p_{\theta}(y^{(h)} | x^{(h)} \oplus x^{(s)}),
    \label{eq:jailbreak-obj}
\end{align}
where $\oplus$ denotes concatenation.
Solving Eq.~(\ref{eq:jailbreak-obj}) typically involves iterative discrete optimization.
In the \textit{heuristic search} phase, algorithms mutate the current suffix to generate a batch of $N$ candidates, $\mathcal{S} = \{s_1, \dots, s_N\}$.
Subsequently, during the \textit{candidate scoring} phase, these candidates are concatenated with the static prompt and the target response to form the batch $\{x^{(h)} \oplus s_i \oplus y^{(h)} \}_{i=1}^N$. This batch is then fed into the model to compute the prediction logits for the target sequence.
Based on the attack-specific scoring objective, the top candidates are selected to seed the next iteration.
Crucially, this cycle repeats for hundreds of epochs, meaning without prefix caching, the static prefix $x^{(h)}$ is re-evaluated $N$ times per step, creating a massive computational workload.

\textbf{Standard Attention and KV Cache.}
The core operator in Transformer-based LLMs~\citep{vaswani2023attentionneed} is the scaled dot-product attention.
For a token position $t$, given query $q_t$ and key-value tensors $(K_{\le t}, V_{\le t})$ from previous and current positions, the attention output $o_t$ is computed as:
\begin{align}
    o_t = \text{Softmax}\left(\frac{q_t K_{\leq t}^\top}{\sqrt{d}}\right) V_{\leq t},
    \label{eq:attention}
\end{align}
To avoid recomputing $K_{\leq t}$ and $V_{\leq t}$ at every generation step, inference systems employ KV Caching to store these tensors in GPU memory.
In a PyTorch-native batched implementation, the KV cache batch dimension must align with the input batch size $N$:
\begin{align}
    \mathcal{M}_{prefix} = \text{Repeat}(K^{(h)}, V^{(h)}, \text{times}=N).
\end{align}
This results in a memory complexity of $\mathcal{O}(N \cdot L_{prefix})$\footnote{Omitting model-dependent constants.}, which strictly limits the search width $N$ and prevents the use of large batch sizes.

\section{Methodology}
\label{sec:method}

In this section, we first analyze redundant prefix KV computation in suffix-based attack evaluation, and then introduce PSKV, a prefix-shared KV-cache framework with layer-wise lazy expansion and suffix-centric alignment.

\subsection{Redundant Prefix computation in Suffix-Based Attack Evaluation}

According to the general jailbreak suffix synthesis process described in Section~\ref{sec:preliminaries}, the main computational burden arises from repeatedly scoring $K\cdot q$ candidate suffixes in each optimization iteration, where $K$ is the number of candidate suffixes in each iteration, and $q$ is the number of new suffixes generated for each candidate.

We observe that each optimization iteration scores up to $Kq$ candidate prompts with the target LLM, and these candidates share the same instruction prefix $x^{(h)}$.
Specifically, as illustrated in Fig.~\ref{fig:general_process}, the suffix jailbreak attack only modifies the suffixes of these candidate prompts but keeps the harmful instruction $x^{(h)}$ as their prefix intact.

The most straightforward implementation, which we term the \textbf{unaccelerated traditional attack method}, treats each of the $K \cdot q$ candidate prompts as independent inputs in every iteration. This approach repeatedly re-computes the KV vectors for the prefix $x^{(h)}$, leading to significant redundant computational overhead.
As a result, when an LLM is performing inference on these prompts in parallel, it will redundantly calculate the same KV vectors for the same prefix $x^{(h)}$ for up to $K \cdot q$ times at each step.

A more advanced baseline would employ a \textbf{standard KV cache} implementation. The KV vectors for the prefix $x^{(h)}$ are computed only once before the iterative synthesis process begins. In a PyTorch-native batched implementation, however, to evaluate all $K \cdot q$ suffix candidates in a single parallel batch, this pre-computed prefix cache must be duplicated $K \cdot q$ times to match the batch dimension of the suffixes, which introduces a severe memory bottleneck. For a batch of $B$ prompts, this problem is exacerbated, requiring $B \cdot K \cdot q$ copies of the caches.

\begin{wraptable}{r}{0.45\textwidth}
\scriptsize
\vspace{-15pt}
    \centering
    \caption{The average length (in tokens) of harmful prompts and targeted texts in HarmBench.}
    \label{tab:prompt_length}
    \vspace{5pt}
    \begin{tabular}{l c c}
    \toprule
    \textbf{Dataset} & \textbf{Harmful Prompt} & \textbf{Targeted Text} \\
    \midrule
    HarmBench & 78.4  & 93.3  \\
    \bottomrule
    \end{tabular}
\vspace{-10pt}
\end{wraptable}

Furthermore, as shown in Table~\ref{tab:prompt_length},
the average token length of instructions in HarmBench is 78.4~\citep{mazeika2024harmbench}, which is significantly larger than the standard jailbreak suffix length in practice (typically set to 20 tokens).
Collectively, these observations suggest that if the KV vectors of the prefix harmful instruction $x^{(h)}$ can be calculated only once and used for the inference of different candidate prompts, then both the \textit{computation} and \textit{memory} costs of suffix jailbreaking can be considerably reduced, as a large amount of redundant intermediate KV vectors no longer need to be calculated.

\subsection{Prefix-Shared KV Cache with Layer-Wise Lazy Expansion}
\label{sec:method:single-suffix}

\begin{wrapfigure}{r}{0.55\textwidth}
    \vspace{-20pt}
  \includegraphics[width=0.55\columnwidth]{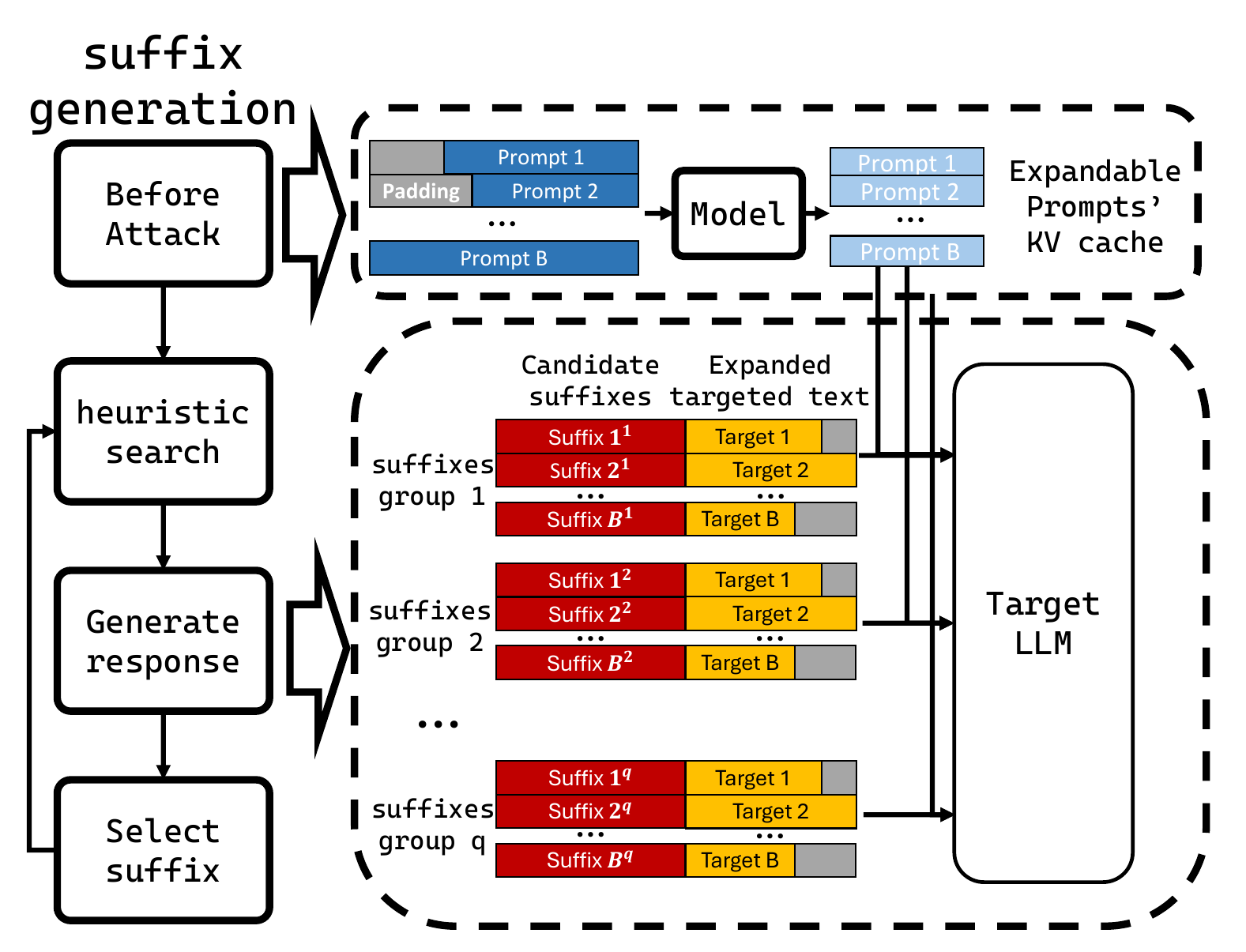}
  \vspace{-15pt}
  \caption{PSKV stores compact prefix KV caches and reuses each prompt's cache across its candidate suffixes, avoiding persistent candidate-expanded prefix caches through layer-wise lazy expansion.}
  \vspace{-20pt}
  \label{fig:ours}
  
\end{wrapfigure}
These observations highlight a clear opportunity for optimization. Based on our previous analysis, we now propose \textit{PSKV}, a method that can substantially accelerate existing suffix jailbreak attacks by reducing redundant computation of prefix KV vectors during the candidate prompt inference process of suffix jailbreaking.

PSKV adopts a prefix-shared, layer-wise lazy expansion strategy. Before the suffix optimization loop starts, PSKV performs a single forward pass on the shared instruction prefix $x^{(h)}$ and stores its KV tensors for all attention layers. During candidate scoring, PSKV does not materialize a fully expanded prefix cache across all layers. Instead, at each transformer layer, it retrieves the compact prefix KV tensors for that layer and temporarily expands them along the candidate dimension inside the attention computation. After the layer output is computed, the expanded tensors are discarded before moving to the next layer.

This design preserves the computational benefit of standard KV caching while avoiding its persistent memory blow-up. The prefix is computed only once, reducing redundant prefix computation from $Kq$ times to one time. At the same time, PSKV stores only one compact prefix-cache set, containing one KV tensor pair per layer,
rather than maintaining $Kq$ candidate-expanded copies. Since the suffix-side forward pass is implemented with standard PyTorch tensor operations, gradients with respect to suffix embeddings remain available for gradient-based attacks. A step-by-step guide for integrating PSKV into suffix-based attack pipelines and implementation details is provided in Appendix~\ref{app:using-pskv}.

\begin{table*}[t]
\centering
\scriptsize
\caption{Unified complexity analysis. $E$: epochs; $N_{cand}$: total candidates per iteration ($K \cdot q$ for single, $B \cdot K \cdot q$ for batch); $N_p, N_s, N_t$: prefix, suffix, and target lengths; $N'_p$: maximum padded prefix length in batched evaluation. $L_{dec} = N_s + N_t$ denotes decoding length.}
\label{tab:complexity}
\begin{tabular}{llcc}
\toprule
\textbf{Scenario} & \textbf{Method} & \textbf{Computational Complexity} & \textbf{Memory Complexity} \\
\midrule
\multirow{3}{*}{\shortstack[l]{Single\\Instruction}} 
 & No Accel. & $\mathcal{O}(E \cdot N_{cand} \cdot (N_p + L_{dec})^2)$ & $\mathcal{O}(N_{cand} \cdot (N_p + L_{dec}))$ \\
 & Standard KV & $\mathcal{O}(N_p^2 + E \cdot N_{cand} \cdot L_{dec} \cdot (2N_p + L_{dec}))$ & $\mathcal{O}(N_{cand} \cdot (\mathbf{N_p} + L_{dec}))$ \\
 & \textbf{PSKV} & $\mathcal{O}(N_p^2 + E \cdot N_{cand} \cdot L_{dec} \cdot (2N_p + L_{dec}))$ & $\mathcal{O}(\mathbf{1} \cdot N_p + N_{cand} \cdot L_{dec})$ \\
\midrule
\multirow{3}{*}{\shortstack[l]{Batched Multi-\\Instruction}} 
 & No Accel. & $\mathcal{O}(E \cdot N_{cand} \cdot (N'_p + L_{dec})^2)$ & $\mathcal{O}(N_{cand} \cdot (N'_p + L_{dec}))$ \\
 & Standard KV & $\mathcal{O}(B \cdot {N'_p}^2 + E \cdot N_{cand} \cdot L_{dec} \cdot (2N'_p + L_{dec}))$ & $\mathcal{O}(N_{cand} \cdot (N'_p + L_{dec}))$ \\
 & \textbf{PSKV} & $\mathcal{O}(B \cdot {N'_p}^2 + E \cdot N_{cand} \cdot L_{dec} \cdot (2N'_p + L_{dec}))$ & $\mathcal{O}(B \cdot N'_p + N_{cand} \cdot L_{dec})$ \\
\bottomrule
\end{tabular}
\vspace{-10pt}
\end{table*}



\subsection{Suffix-Centric Alignment for Batched Multi-Instruction Evaluation}
\label{sec:method:batch-suffix}

In real-world benchmarking, evaluation requires optimizing suffixes for a batch of $B$ distinct instruction-response pairs $\{(x_i^{(h)}, y_i^{(h)})\}_{i=1}^B$.
A practical challenge arises from the variable token lengths of these pairs, which typically leads to ragged tensors that hinder efficient parallelization.
\begin{wrapfigure}{r}{0.6\textwidth}
    
    \includegraphics[width=0.6\columnwidth]{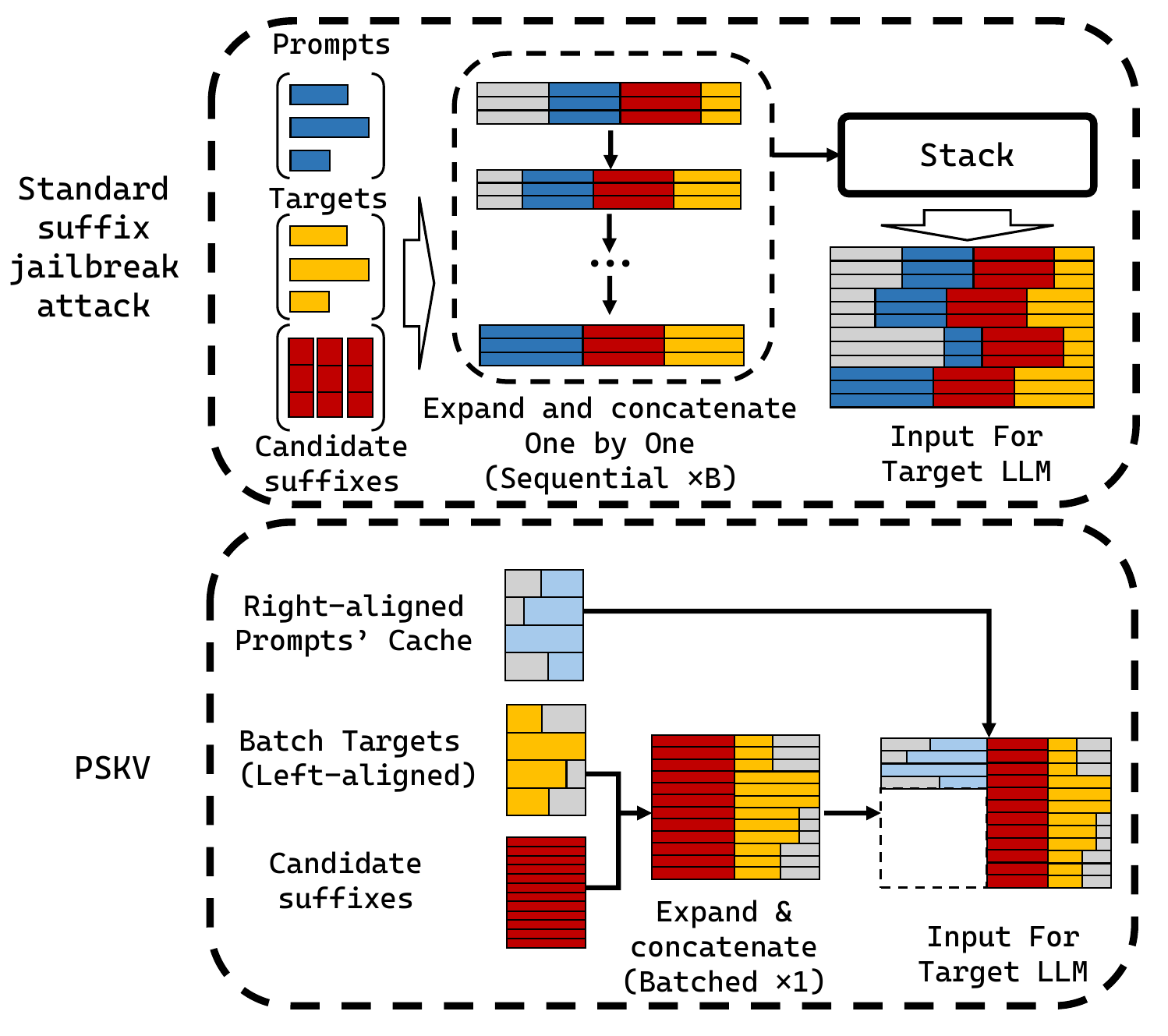}
  \vspace{-10pt}
    \caption{Alignment strategies for batched suffix attacks. The standard method’s misaligned structure prevents prompt KV cache reuse. PSKV uses suffix-centric alignment, which enables each prompt's KV cache to be reused across its candidate suffixes and allows suffix-side operations to be processed as dense tensor operations.}
  \label{fig:suffix_align}
  \vspace{-30pt}
\end{wrapfigure}


For batched multi-instruction evaluation, PSKV further uses suffix-centric alignment. We left-pad instructions and right-pad target responses so that all suffix tokens start at the same column index. This produces a dense suffix tensor of shape $(B\cdot K\cdot q)\times N_s$, enabling candidate substitution, gradient extraction, and scoring to be implemented as vectorized tensor operations. Meanwhile, PSKV computes one prefix cache per instruction and reuses each cache across the corresponding $Kq$ candidate suffixes, reducing the prefix-cache memory term from $\mathcal{O}(B K q \cdot N'_p)$ to $\mathcal{O}(B\cdot N'_p)$.




\subsection{Complexity Analysis}
\label{sec:method:complexity}

We summarize the computational and resident cache memory of \textbf{No Acceleration}, \textbf{Standard KV Cache}, and \textbf{PSKV} in Table~\ref{tab:complexity}. Let $E$ denote the number of synthesis iterations, $N_s$ the suffix length, $N_t$ the target length, and $L_{dec}=N_s+N_t$ the suffix-side decoding length. For single-instruction evaluation, $N_{cand}=Kq$ denotes the number of candidate suffixes per iteration. For batched multi-instruction evaluation, $N_{cand}=BKq$.

Standard KV Cache and PSKV have the same asymptotic computational complexity because both compute the shared prefix once and reuse its KV tensors during candidate scoring. Their difference lies in resident cache memory. Standard KV Cache must expand the prefix KV tensors to the full candidate batch, resulting in a prefix-cache memory term proportional to $\mathcal{O}(N_{cand}N_p)$ for single-instruction evaluation and $\mathcal{O}(N_{cand}N'_p)$ for batched multi-instruction evaluation. In contrast, PSKV stores only one prefix cache for the single-instruction case, or one prefix cache per instruction in the batched case. Therefore, the persistent prefix-cache memory is reduced to $\mathcal{O}(N_p)$ and $\mathcal{O}(B N'_p)$, respectively, while the suffix-side memory remains proportional to $\mathcal{O}(N_{cand}L_{dec})$. Detailed derivations are provided in Appendix~\ref{appendix:complexity} and~\ref{appendix:complexity-batch}. For PSKV, the candidate-expanded prefix KV tensors exist only as transient per-layer workspace and are not stored persistently across layers\footnote{The implementation details are provided in Appendix~\ref{app:lazy-expansion-overhead}}.

\section{Experiments}
\label{sec:experiments}
In this section, we empirically explore how our proposed PSKV method can reduce the computation and memory costs of suffix-based jailbreak attacks while preserving their attack effectiveness.

\subsection{Experimental Setup}

\textbf{Safety dataset.}
We use the first $50$ examples from the test set of \textbf{HarmBench}~\citep{mazeika2024harmbench} to form the safety dataset for the suffix jailbreaking evaluations.
Each sample from this safety dataset consists of a harmful instruction and a targeted harmful response.

\textbf{Targeted Models.}
We empirically validated PSKV using six suffix attacks on five target LLMs, which are:
\textbf{Vicuna-7B-v1.5}~\citep{lmsys2023vicuna}, \textbf{Llama-2-7B-Chat}~\citep{touvron2023llama}, \textbf{Llama-3-8B-Instruct}~\citep{grattafiori2024llama3}, \textbf{Mistral-7B-Instruct-v0.3}~\citep{jiang2023mistral7b}, and \textbf{Qwen2.5-7B-Instruct}~\citep{qwen2025qwen25}.

\textbf{Suffix jailbreak attacks.}
We benchmark how PSKV improves the efficiency of evaluating six common suffix-based jailbreak methods from two families.
Specifically, \textbf{GCG}~\citep{zou2023universal}, \textbf{GCQ}~\citep{hayase2024query}, \textbf{Zhu's AutoDAN}~\citep{zhu2023autodan} (abbreviated as \textbf{AutoDAN} hereafter), and \textbf{BEAST}~\citep{sadasivan2024fast} are from the family of optimization-based attacks, which aim to search for jailbreak suffixes using various discrete optimization techniques.
\textbf{AmpleGCG}~\citep{liao2024amplegcg} and \textbf{AdvPrompter}~\citep{paulus2024advprompter} are from the family of model-based attacks, which aim to train LLM-based generators to directly generate jailbreak suffixes for given harmful instructions.
Detailed hyperparameters for all attacks are provided in Appendix~\ref{app:jailbreak-impl}. Unless otherwise specified, all experiments are run on two NVIDIA A100 GPUs. We report wall-clock time and aggregate peak GPU memory across the two GPUs.

\textbf{Evaluation metrics.}
For the attack performance evaluation, we report the \textbf{Attack Success Rate (ASR)}, defined as the percentage of successful jailbreak attacks on the safety dataset. Please refer to Appendix~\ref{app:jailbreak-impl} for details. For attack efficiency evaluation, we adopt memory consumption in gigabytes (GB) and execution time in minutes as the metrics.

\textbf{Baseline acceleration methods.}
Our experiments evaluate the effectiveness of our proposed \textit{PSKV} by comparing it against two alternatives: a baseline implementation where \textbf{no specific acceleration method is deployed} (denoted as \textbf{None}), and an implementation using a \textbf{Standard KV Cache}. 

\textbf{None} represents an unaccelerated, albeit batched, approach. In each iteration, it constructs $K \cdot q$ full candidate prompts by concatenating the prefix $x^{(h)}$ with each suffix, then performs a single, large batched forward pass on these sequences. This method redundantly calculates the prefix KV vectors for every candidate in every iteration.
\textbf{Standard KV Cache} implements the intuitive optimization of pre-computing the prefix cache. It first performs a single forward pass on the prefix $x^{(h)}$ to obtain its KV cache. Then, to evaluate the $K \cdot q$ suffix candidates in a parallel batch, it physically replicates this prefix cache $K \cdot q$ times on GPU to match the batch dimension of the suffixes.
 
\subsection{Results}

\begin{table*}[thb]
\centering

\scriptsize
\caption{
Attack Success Rates (ASRs, \%) of suffix jailbreak attacks with and without PSKV. Values are reported as mean $\pm$ standard deviation over three runs.
}
\label{tab:asr}
\begin{tabular}{l l c c c c c c}
\toprule
 \textbf{Model} & \textbf{Cache} & \textbf{GCG} & \textbf{GCQ} & \textbf{AutoDAN} & \textbf{BEAST} & \textbf{AmpleGCG} & \textbf{AdvPrompter} \\
\midrule

 \multirow{2}{*}{Vicuna-7B}
& None & $98.0 \pm 2.0$ & $90.0 \pm 3.5$ & $54.7 \pm 3.1$ & $94.0 \pm 2.0$ & $94.7 \pm 3.1$ & $97.3 \pm 1.2$\\
 & PSKV& $96.0 \pm 2.0$ & $93.3 \pm 2.3$ & $53.3 \pm 5.0$ & $95.3 \pm 4.2$ & $93.3 \pm 2.3$ & $96.0 \pm 2.0$\\
 \midrule
 \multirow{2}{*}{Llama-2-7B}
& None & $64.0 \pm 3.5$ & $18.0 \pm 0.0$ & $24.7 \pm 1.2$ & $39.3 \pm 3.1$ & $93.3 \pm 2.3$ & $91.3 \pm 4.2$\\
 & PSKV& $66.7 \pm 1.2$ & $17.3 \pm 1.2$ & $25.3 \pm 3.1$ & $42.7 \pm 3.1$ & $92.7 \pm 4.2$ & $92.0 \pm 5.3$\\
 \midrule
 \multirow{2}{*}{Llama-3-8B}
& None & $62.7 \pm 2.3$ & $25.3 \pm 1.2$ & $28.0 \pm 4.0$ & $48.7 \pm 3.1$ & $70.7 \pm 2.3$ & $74.7 \pm 3.1$\\
 & PSKV & $65.3 \pm 9.5$ & $26.0 \pm 0.0$ & $28.0 \pm 2.0$ & $50.7 \pm 4.2$ & $72.0 \pm 3.5$ & $77.3 \pm 4.2$\\
 \midrule
 \multirow{2}{*}{Mistral-7B}
& None & $93.3 \pm 4.2$ & $91.3 \pm 3.1$ & $82.7 \pm 4.2$ & $85.3 \pm 3.1$ & $86.0 \pm 4.0$ & $92.7 \pm 2.3$\\
 & PSKV& $94.0 \pm 4.0$ & $92.0 \pm 5.3$ & $81.3 \pm 3.1$ & $86.7 \pm 3.1$ & $88.7 \pm 4.2$ & $93.3 \pm 2.3$\\
 \midrule
 \multirow{2}{*}{Qwen-2.5-7B}
& None & $86.0 \pm 2.0$ & $86.7 \pm 3.1$ & $55.3 \pm 4.2$ & $89.3 \pm 3.1$ & $82.0 \pm 3.5$ & $92.7 \pm 4.2$\\
 & PSKV& $89.3 \pm 2.3$ & $89.3 \pm 2.3$ & $55.3 \pm 5.0$ & $87.3 \pm 1.2$ & $86.7 \pm 2.3$ & $93.3 \pm 1.2$\\

\bottomrule
\end{tabular}%
\vspace{-10pt}
\end{table*}

\begin{table*}[t]
\scriptsize
\centering
\caption{
Time usage (min) for (1) optimization-based suffix synthesis and (2) model-based training sample generation. Parentheses ($\times$) indicate speed-up factors vs. the unaccelerated baseline. \textbf{OOM}: Out Of Memory.
}
\label{tab:time-usage}
\begin{tabular}{l l c c c c c c}
\toprule
\multirow{2}{*}{\textbf{Model}} & \multirow{2}{*}{\textbf{Acceleration}}
& \multicolumn{4}{c}{\textbf{Optimization-based Attacks}}
& \multicolumn{2}{c}{\textbf{Model-based Attacks}} \\
\cmidrule(lr){3-6}
\cmidrule(lr){7-8}
& & \textbf{GCG} & \textbf{GCQ} & \textbf{AutoDAN} & \textbf{BEAST} & \textbf{AmpleGCG} & \textbf{AdvPrompter} \\
\midrule
\multirow{3}{*}{Vicuna-7B}
& None & 228 \scriptsize($\times$ 1.00)& 247 \scriptsize($\times$ 1.00) & 88 \scriptsize($\times$ 1.00)& 23 \scriptsize($\times$ 1.00)& 349 \scriptsize($\times$ 1.00)& 462 \scriptsize($\times$ 1.00)\\
& Standard KV & 147 \scriptsize($\times$ 1.55)& OOM& OOM& OOM& 239 \scriptsize($\times$ 1.46)& 160 \scriptsize($\times$ 2.89)\\
& PSKV (Ours) & 146 \scriptsize($\times$ 1.56)& 148 \scriptsize($\times$ 1.67)& 50 \scriptsize($\times$ 1.76)& 15 \scriptsize($\times$ 1.53)& 253 \scriptsize($\times$ 1.38)& 199 \scriptsize($\times$ 2.32)\\
\midrule
\multirow{3}{*}{Llama-2-7B}
& None & 221 \scriptsize($\times$ 1.00)& 210 \scriptsize($\times$ 1.00)& 86 \scriptsize($\times$ 1.00)& 26
\scriptsize($\times$ 1.00)& 349 \scriptsize($\times$ 1.00)& 513 \scriptsize($\times$ 1.00)\\
& Standard KV & 142 \scriptsize($\times$ 1.56)& OOM& OOM& OOM& 242 \scriptsize($\times$ 1.44)& 344 \scriptsize($\times$ 1.49)\\
& PSKV (Ours) & 142 \scriptsize($\times$ 1.56)& 145 \scriptsize($\times$ 1.45)& 48 \scriptsize($\times$ 1.79)& 14 \scriptsize($\times$ 1.86)& 227 \scriptsize($\times$ 1.54)& 275 \scriptsize($\times$ 1.87)\\
\midrule
\multirow{3}{*}{Llama-3-8B} 
& None & 288 \scriptsize($\times$ 1.00)& 221 \scriptsize($\times$ 1.00)& 88 \scriptsize($\times$ 1.00)& 24 \scriptsize($\times$ 1.00)& 329 \scriptsize($\times$ 1.00)& 577 \scriptsize($\times$ 1.00)\\
& Standard KV & 151 \scriptsize($\times$ 1.91)& 141 \scriptsize($\times$ 1.57)& 48 \scriptsize($\times$ 1.83)& OOM& 240 \scriptsize($\times$ 1.37)& 282 \scriptsize($\times$ 2.05)\\
& PSKV (Ours) & 141 \scriptsize($\times$ 2.04)& 138 \scriptsize($\times$ 1.60)& 45 \scriptsize($\times$ 1.96)& 14 \scriptsize($\times$ 1.71)& 242 \scriptsize($\times$ 1.36)& 307 \scriptsize($\times$ 1.88)\\
\midrule
\multirow{3}{*}{Mistral-7B} 
& None & 211 \scriptsize($\times$ 1.00)& 203 \scriptsize($\times$ 1.00)& 81 \scriptsize($\times$ 1.00)& 25 \scriptsize($\times$ 1.00)& 320 \scriptsize($\times$ 1.00)& 96 \scriptsize($\times$ 1.00)\\
& Standard KV & 138 \scriptsize($\times$ 1.53)& 126 \scriptsize($\times$ 1.61)& 40 \scriptsize($\times$ 2.03)& OOM& 229 \scriptsize($\times$ 1.40)& 80 \scriptsize($\times$ 1.20)\\
& PSKV (Ours) & 131 \scriptsize($\times$ 1.61)& 122 \scriptsize($\times$ 1.66)& 40 \scriptsize($\times$ 2.03)& 14 \scriptsize($\times$ 1.79)& 230 \scriptsize($\times$ 1.39)& 74 \scriptsize($\times$ 1.30)\\
\midrule
\multirow{3}{*}{Qwen2.5-7B} 
& None & 250 \scriptsize($\times$ 1.00)& 242 \scriptsize($\times$ 1.00)& 84 \scriptsize($\times$ 1.00)& 26 \scriptsize($\times$ 1.00)& 295 \scriptsize($\times$ 1.00)& 102 \scriptsize($\times$ 1.00)\\
& Standard KV & 134 \scriptsize($\times$ 1.87)& 143 \scriptsize($\times$ 1.69)& 44 \scriptsize($\times$ 1.91)& OOM& 223 \scriptsize($\times$ 1.32)& 81 \scriptsize($\times$ 1.26)\\
& PSKV (Ours) & 135 \scriptsize($\times$ 1.85)& 128 \scriptsize($\times$ 1.89)& 42 \scriptsize($\times$ 2.00)& 15 \scriptsize($\times$ 1.73)& 237 \scriptsize($\times$ 1.24)& 76 \scriptsize($\times$ 1.34)\\
\bottomrule
\end{tabular}%
\vspace{-10pt}
\end{table*}

\textbf{Attack effectiveness.}
According to the design of PSKV described in Section~\ref{sec:method}, PSKV is a performance-neutral acceleration method that does not affect the attack performance of accelerated suffix jailbreak attacks. This is because PSKV only avoids redundant prefix-KV computation and persistent candidate-expanded cache storage; it does not change the attack algorithm or numerical precision.

Table~\ref{tab:asr} reports three-run ASR mean and standard deviation, showing that PSKV preserves attack effectiveness within the natural run-to-run variation of the underlying attacks.
To further verify that the observed ASR fluctuations are not caused by PSKV, we conduct two additional checks in Appendix~\ref{app:asr-stability}: (1) a paired 10-seed GCG comparison between PSKV and No Cache on the same data, which yields no statistically significant ASR difference; and (2) a logit-equivalence test showing that PSKV produces identical logits to Standard KV Cache on identical inputs under the same precision setting.
Together, these results confirm that PSKV preserves the behavior of the underlying attacks while only accelerating shared-prefix inference computation.

\textbf{PSKV considerably accelerates suffix jailbreaking.}
Table~\ref{tab:time-usage} details the time consumption of varying acceleration strategies. For attacks whose runtime is dominated by repeated candidate evaluation (including GCG, GCQ, BEAST, and AmpleGCG), PSKV consistently achieves a speed-up ratio of approximately $1.4\times$ to $1.6\times$ compared to the no-acceleration baseline. Notably, this acceleration becomes even more pronounced for Zhu's AutoDAN, where the speed-up ratio rises to a range of $\sim1.8\times$ to $1.95\times$ (e.g., reducing time from 88 to 45 minutes on Llama-3-8B). This performance gain is attributed to the mechanism of KV cache acceleration: it reduces computational redundancy by storing states. Since AutoDAN requires more inference steps per iteration (due to hierarchical genetic algorithm), there is a larger proportion of redundant computation for PSKV to eliminate.

Conversely, for AdvPrompter, the speed-up ratio decreases to approximately $1.3\times$. This is because its training loop includes both target-model
candidate evaluation and AdvPrompter model fine-tuning. While PSKV accelerates the inference phase, the training overhead remains constant, diluting the overall proportion of time saved. Furthermore, compared to the Standard KV Cache, our method maintains a comparable speed without incurring time penalties (in some cases, such as Llama-3 AdvPrompter, it is even faster). This indicates that the overhead caused by our temporary KV tensor expansion is negligible relative to the total inference time.

Across target models, runtime differences are influenced by model architecture, tokenizer behavior, and attack-specific overheads, but the relative acceleration trend of PSKV remains consistent.\footnote{We also observe that AdvPrompter requires significantly more time on Llama-series models compared to Mistral and Qwen. For explanations, please refer to App.~\ref{app:advPrompter}.} Since the sequence lengths are relatively short (avg. $\sim$190 tokens), the bottleneck is dominated by repeated forward computation rather than by request-scheduling or serving-throughput limitations.

\begin{table*}[t]
\scriptsize
\centering
\caption{
Memory consumption (GB) for (1) optimization-based suffix synthesis and (2) model-based training sample generation. \textbf{OOM}: Out Of Memory. ($\times$): $\text{Mem}_{\mathrm{StandardKV}} / \text{Mem}_{\mathrm{method}}$; ($-$) denotes baseline OOM.
}
\label{tab:memory-usage}
\begin{tabular}{l l c c c c c c}
\toprule
\multirow{2}{*}{\textbf{Model}} & \multirow{2}{*}{\textbf{Acceleration}}
& \multicolumn{4}{c}{\textbf{Optimization-based Attacks}}
& \multicolumn{2}{c}{\textbf{Model-based Attacks}} \\
\cmidrule(lr){3-6}
\cmidrule(lr){7-8}
& & \textbf{GCG} & \textbf{GCQ} & \textbf{AutoDAN} & \textbf{BEAST} & \textbf{AmpleGCG} & \textbf{AdvPrompter} \\
\midrule
\multirow{3}{*}{Vicuna-7B}
& None
& 30 \scriptsize($\times$ 3.35)& 67 \scriptsize(-)& 90 \scriptsize(-)& 144 \scriptsize(-)& 146 \scriptsize($\times$ 1.02)& 125 \scriptsize($\times$ 1.25)\\
& Standard KV
& 100 \scriptsize($\times$ 1.00)& OOM& OOM& OOM& 148 \scriptsize($\times$ 1.00)& 155 \scriptsize($\times$ 1.00)\\
& PSKV (Ours)
& 27 \scriptsize($\times$ 3.68)& 56 \scriptsize(-)& 53 \scriptsize(-)& 92 \scriptsize(-)& 63 \scriptsize($\times$ 2.36)& 67 \scriptsize($\times$ 2.32)\\
\midrule
\multirow{3}{*}{Llama-2-7B}
& None
& 35 \scriptsize($\times$ 2.84)& 63 \scriptsize(-)& 102 \scriptsize(-)& 112 \scriptsize(-)& 145 \scriptsize($\times$ 1.02)& 73 \scriptsize($\times$ 2.09)\\
& Standard KV
& 100 \scriptsize($\times$ 1.00)& OOM& OOM& OOM& 148 \scriptsize($\times$ 1.00)& 152 \scriptsize($\times$ 1.00)\\
& PSKV (Ours)
& 30 \scriptsize($\times$ 3.39)& 50 \scriptsize(-)& 51 \scriptsize(-)& 81 \scriptsize(-)& 58 \scriptsize($\times$ 2.55)& 63 \scriptsize($\times$ 2.40)\\
\midrule
\multirow{3}{*}{Llama-3-8B}
& None
& 56 \scriptsize($\times$ 1.05)& 118 \scriptsize($\times$ 1.06)& 124 \scriptsize($\times$ 1.10)& 142 \scriptsize(-)& 128 \scriptsize($\times$ 0.88)& 117 \scriptsize($\times$ 1.27)\\
& Standard KV
& 58 \scriptsize($\times$ 1.00)& 125 \scriptsize($\times$ 1.00)& 136 \scriptsize($\times$ 1.00)& OOM& 112 \scriptsize($\times$ 1.00)& 149 \scriptsize($\times$ 1.00)\\
& PSKV (Ours)
& 41 \scriptsize($\times$ 1.43)& 85 \scriptsize($\times$ 1.47)& 77 \scriptsize($\times$ 1.77)& 110 \scriptsize(-)& 91 \scriptsize($\times$ 1.24)& 126 \scriptsize($\times$ 1.18)\\
\midrule
\multirow{3}{*}{Mistral-7B}
& None
& 34 \scriptsize($\times$ 1.33)& 54 \scriptsize($\times$ 1.56)& 85 \scriptsize($\times$ 1.35)& 144 \scriptsize(-)& 86 \scriptsize($\times$ 0.93)& 54 \scriptsize($\times$ 1.69)\\
& Standard KV
& 46 \scriptsize($\times$ 1.00)& 84 \scriptsize($\times$ 1.00)& 114 \scriptsize($\times$ 1.00)& OOM& 80 \scriptsize($\times$ 1.00)& 90 \scriptsize($\times$ 1.00)\\
& PSKV (Ours)
& 28 \scriptsize($\times$ 1.62)& 44 \scriptsize($\times$ 1.91)& 44 \scriptsize($\times$ 2.59)& 120 \scriptsize(-)& 58 \scriptsize($\times$ 1.37)& 66 \scriptsize($\times$ 1.37)\\
\midrule
\multirow{3}{*}{Qwen2.5-7B}
& None
& 66 \scriptsize($\times$ 0.81)& 122 \scriptsize($\times$ 0.94)& 141 \scriptsize($\times$ 0.89)& 139 \scriptsize(-)& 121 \scriptsize($\times$ 0.93)& 127 \scriptsize($\times$ 1.15)\\
& Standard KV
& 54 \scriptsize($\times$ 1.00)& 115 \scriptsize($\times$ 1.00)& 125 \scriptsize($\times$ 1.00)& OOM& 112 \scriptsize($\times$ 1.00)& 146 \scriptsize($\times$ 1.00)\\
& PSKV (Ours)
& 42 \scriptsize($\times$ 1.28)& 94 \scriptsize($\times$ 1.22)& 82 \scriptsize($\times$ 1.53)& 115 \scriptsize(-)& 94 \scriptsize($\times$ 1.19)& 138 \scriptsize($\times$ 1.06)\\
\bottomrule

\end{tabular}%
\vspace{-10pt}
\end{table*}

\paragraph{\textbf{Memory Efficiency Analysis of PSKV.}}

Table~\ref{tab:memory-usage} reveals that PSKV consistently reduces memory relative to Standard KV Cache and often uses less memory than the unaccelerated batched baseline by avoiding persistent candidate-expanded prefix KV caches. Unlike Standard KV Cache, which suffers from memory explosion under large beam widths by maintaining a fully expanded state, PSKV leverages layer-wise inference to dynamically expand and immediately discard the KV view for the current layer, keeping the resident footprint minimal.

Memory savings vary by attack dynamics: for gradient-based attacks (GCG) where the KV cache dominates memory usage, PSKV achieves high compression; conversely, while massive intermediate tensors in beam-search methods (GCQ, BEAST) lower the relative reduction ratio, PSKV remains vital for preventing the OOM errors frequent in Standard Cache. Similarly, PSKV mitigates cumulative cache growth in AutoDAN, though in dual-model setups like AdvPrompter, the increased non-KV memory burden can marginally offset these savings.

Architecturally, the efficiency gap is most pronounced with Multi-Head Attention (MHA) models (Vicuna-7B, Llama-2-7B) where PSKV eliminates the redundancy of storing unique pairs for every head.\footnote{See Appendix~\ref{app:memory_analysis} for architectural details.} In contrast, models utilizing Grouped-Query Attention (GQA) like Llama-3, Mistral,\footnote{Mistral's sliding window does not impact these results as inputs fit within the window.} and Qwen2.5 inherently reduce the baseline KV size, lowering PSKV's theoretical compression ceiling. Furthermore, the larger vocabularies in these models generate massive intermediate tensors (logits, IDs) that further dilute the KV cache's relative memory proportion, rendering relative savings less apparent. See Appendix~\ref{app:memory_analysis} for a more detailed analysis of memory consumption.

\paragraph{\textbf{Scalability Analysis.}}
Table~\ref{tab:iter_perf} highlights PSKV's efficient memory scalability. Remarkably, the total peak memory consumption remains virtually constant despite a 4$\times$ increase in search width (e.g., GCG stabilizes at $\approx$40.8 GB across all widths). This stability arises because the peak memory footprint is dominated by substantial intermediate tensors required for gradient and loss computation, rather than the inference cache itself. Since PSKV enables efficient prefix sharing, the marginal memory cost of expanding the search width is negligible and remains entirely masked beneath the high-water mark set by these non-cache components.\footnote{See Appendix~\ref{app:longer-contexts} for the proof.} Consequently, the amortized memory cost per candidate drops inversely with the width, demonstrating that PSKV enables high-width parallel search without raising the hardware memory barrier. We further evaluate PSKV under longer suffixes and longer instruction prefixes in Appendix~\ref{app:longer-contexts}. 
The results and memory snapshots show that PSKV avoids both the full-sequence forward peaks of No Cache and the persistent expanded-cache plateau of Standard KV Cache, making its memory advantage increasingly pronounced for longer-prefix workloads. 
\begin{wraptable}{r}{0.6\textwidth}
\vspace{20pt}
\centering
\scriptsize
\caption{Scalability of iterative/search-based attacks with PSKV on HarmBench using a Llama-3-8B target.}
\begin{tabular}{l l cccc c}
\toprule
\multirow{2}{*}{\textbf{Attack}} & \multirow{2}{*}{\textbf{Width}} & \multicolumn{2}{c}{\textbf{Total Cost}} & \multicolumn{2}{c}{\textbf{Cost/width}} & \multirow{2}{*}{\textbf{ASR (\%)}} \\
\cmidrule(lr){3-4} \cmidrule(lr){5-6}
& & Time & Mem. & Time & Mem. & \\
& & (min) & (GB) & (min) & (GB) & \\
\midrule
\multirow{3}{*}{GCG} 
& 16 & 40  & 40.83 & 2.50 & 2.55 & 50 \\
& 32 & 73  & 40.83 & 2.28 & 1.28 & 54 \\
& 64 & 141 & 40.84 & 2.20 & 0.64 & 62 \\
\midrule
\multirow{3}{*}{GCQ} 
& 16 & 34  & 85.06 & 2.13 & 5.32 & 26 \\
& 32 & 70  & 85.07 & 2.19 & 2.66 & 26 \\
& 64 & 138 & 85.07 & 2.16 & 1.33 & 26 \\
\midrule
\multirow{3}{*}{BEAST} 
& 5  & 5   & 110.27 & 1.00 & 22.05 & 40 \\
& 10 & 9   & 110.29 & 0.90 & 11.03 & 50 \\
& 15 & 14  & 110.29 & 0.93 & 7.35  & 54 \\
\bottomrule
\end{tabular}
\label{tab:iter_perf}
\vspace{-20pt}
\end{wraptable}

\begin{wraptable}{r}{0.6\textwidth}

\centering
\small
\scriptsize
\caption{Efficiency of gradient-free BEAST attack on specialized inference frameworks (SGLang and vLLM).}
\begin{tabular}{ll ccccc}
\toprule
\multirow{2}{*}{\textbf{Framework}}&  \multirow{2}{*}{\textbf{Cost}}&\multicolumn{5}{c}{\textbf{Model}}\\
 \cmidrule(lr){3-7}&  &vicuna& llama2& llama3&mistral& Qwen2.5\\
\midrule
\multirow{2}{*}{\textbf{sglang}}&  Time&8:49& 9:25& 6:22&7:00& 6:12\\
&  Mem.&13.42& 12.87& 14.96&13.50& 14.23\\
\midrule
\multirow{2}{*}{\textbf{vLLM}}&  Time& 16:02 & 16:50& 17:58&16:27& 15:51\\
&  Mem.& 12.55& 12.57 & 14.96 &12.73& 13.52\\
\bottomrule
\end{tabular}
\label{tab:vllm_sglang}
\end{wraptable}

\paragraph{Comparisons with State-of-the-Art Inference Systems.}

We include vLLM~\citep{kwon2023efficient} and SGLang~\citep{zheng2024sglang} in our evaluation to benchmark our method against comprehensively optimized inference systems used in production. While these frameworks utilize advanced memory management techniques like PagedAttention, they lack the gradient support required for white-box optimization (e.g., GCG, AdvPrompter). However, for the gradient-free \textbf{BEAST} attack, they serve as critical baselines. By comparing against these highly specialized engines, we evaluate how close PSKV can come to specialized inference engines while retaining PyTorch autograd support.

Table~\ref{tab:vllm_sglang} benchmarks the performance of the BEAST attack on specialized inference frameworks. Although vLLM and SGLang exhibit impressive efficiency through their full-stack optimized kernels, their design explicitly sacrifices gradient computation, making them unsuitable for white-box optimization tasks. PSKV runs in the same order of magnitude as specialized inference engines on BEAST, while retaining compatibility with PyTorch-native attack pipelines and gradient-based methods.
Furthermore, while vLLM and SGLang are restricted to gradient-free attacks, PSKV serves as a universal plug-and-play accelerator across the entire spectrum of suffix jailbreak methodologies, bridging the gap between high-performance inference and flexible adversarial optimization.

\section{Conclusion}
\label{sec:conclusion}
Suffix-based jailbreak evaluation often relies on heuristic search procedures that require a vast number of resource-intensive LLM forward passes, creating a significant barrier to efficient security evaluation. Our work is motivated by the key observation: for a single harmful instruction, numerous candidate suffixes are generated, yet standard methods create multiple redundant copies of the prompt's computational state, leading to substantial memory overhead. To address this inefficiency, we propose the \textit{Prefix-Shared KV Cache (PSKV)} framework. The core principle of PSKV is to maintain only one compact shared KV cache for each harmful instruction, which is reused across candidate suffixes through layer-wise lazy expansion. To enable this sharing mechanism at scale for batches of diverse prompts, we also introduce a suffix-centric alignment strategy. We empirically validated PSKV across six suffix attacks on five LLMs, demonstrating an average reduction of 40\% in time and 50\% in memory relative to the corresponding baselines, while preserving the original attack success rate. We discuss limitations of PSKV in Appendix~\ref{app:limitation}. Because suffix jailbreak acceleration is dual-use, we discuss responsible release and ethical considerations in Appendix~\ref{app:responsible-release}.

\bibliographystyle{plainnat}
\bibliography{PSKV}

\appendix
\section{Implementation Details of Jailbreak Attacks}
\label{app:jailbreak-impl}

\paragraph{Attack Methods.} Our experiments adopt six suffix-based jailbreak attacks, including four optimization-based methods: GCG~\citep{zou2023universal}, BEAST~\citep{sadasivan2024fast}, GCQ~\citep{hayase2024query}, and Zhu's AutoDAN~\citep{zhu2023autodan}, as well as two model-based attacks: AmpleGCG~\citep{liao2024amplegcg} and AdvPrompter~\citep{paulus2024advprompter}.
We re-implemented all six attacks in a unified PyTorch framework to support PSKV integration and efficient batched evaluation. Unless otherwise specified, we fix the jailbreak suffix length to $20$.
The hyperparameters for each attack are configured as follows:

\begin{itemize}
\item
\textbf{GCG:}
Following Algorithm 1 in \citet{zou2023universal} and its \href{https://github.com/llm-attacks/llm-attacks}{official repository}, we set: iterations = 500, top-k = 256, batch size (search width) = 64, prompt batch size = 16, and search width batch size = 16.

\item
\textbf{GCQ:}
Following Algorithm 1 in \citet{hayase2024query}, we set: iterations = 500, top-k = 256, proxy batch size and query batch size = 15, buffer size (search width) = 64, and batch size for prompts and buffer = 16.

\item
\textbf{BEAST:}
Following Algorithm 1 in \citet{sadasivan2024fast} and its \href{https://github.com/vinusankars/BEAST}{official repository}, we set: both beam-search parameters = 15, and beam-search batch size = 15.

\item
\textbf{Zhu's AutoDAN:}
Following Algorithms 1 and 2 in \citet{zhu2023autodan}, we set: iterations = 3, objective weights = (10, 100), top-$B$ = 256, and temperature = 2.

\item
\textbf{AmpleGCG:}
Following Algorithm 1 in \citet{liao2024amplegcg} and its \href{https://github.com/OSU-NLP-Group/AmpleGCG/tree/main}{official repository}, we set: steps = 500, batch size (search width) = 64, and prompt batch size = 16.

\item
\textbf{AdvPrompter:}
Following Algorithm 1 in \citet{paulus2024advprompter} and its \href{https://github.com/facebookresearch/advprompter}{official repository}, we set: epochs = 10, candidate set size = 256, prompt batch size = 8, temperature = 0.6, penalty parameter = 1.2, beam set size = 64, and beam number = 4.
\end{itemize}

\paragraph{Attack Success Rate(ASR).} For each sample from the safety dataset, we perform a jailbreak attack to synthesize a jailbreak suffix and use it to induce the targeted LLMs to generate $10$ responses.
An LLM-based judger~\footnote{\url{https://huggingface.co/cais/HarmBench-Llama-2-13b-cls}} from \citet{mazeika2024harmbench} is then used to evaluate whether a response contains harmful information.
An attack is determined to be successful if any of the $10$ responses is determined to be harmful.

\section{Proof: time and memory complexity}
\subsection{Complexity Analysis for Single Prompt}
\label{appendix:complexity}

In this section, we provide the derivations for the computational ($C$) and memory ($M$) complexities presented in Section~\ref{sec:method:single-suffix}. We let $E$ be the number of synthesis iterations; $N_p, N_s, N_t$ be the token lengths of the prefix, suffix, and target, respectively; and $K \cdot q$ be the number of candidates evaluated per iteration. Our analysis abstracts away model-specific constants (e.g., hidden dimension, number of layers) as they are common factors across all methods. The core assumptions are that the attention mechanism has a computational cost of $\mathcal{O}(L^2)$ for a sequence of length $L$, and the corresponding KV cache has a memory footprint of $\mathcal{O}(L)$.

\subsubsection{Method without Acceleration}

\paragraph{Computational Complexity.} This method recomputes the entire sequence of length $N_p + N_s + N_t$ for each of the $K \cdot q$ candidates in every one of the $E$ iterations. The total complexity is the product of the number of operations and the cost per operation:
\begin{align}
    C_{\text{NoAccel}} &= E \cdot (K \cdot q) \cdot \mathcal{O}((N_p + N_s + N_t)^2) \nonumber \\
    &= \mathcal{O}(E \cdot (K \cdot q) \cdot (N_p + N_s + N_t)^2).
\end{align}

\paragraph{Memory Complexity.} The peak memory is determined by the batched inference step, which must hold the KV caches for all $K \cdot q$ candidates simultaneously. The memory for each cache is proportional to the full sequence length.
\begin{align}
    M_{\text{NoAccel}} &= (K \cdot q) \cdot \mathcal{O}(N_p + N_s + N_t) \nonumber \\
    &= \mathcal{O}((K \cdot q) \cdot (N_p + N_s + N_t)).
\end{align}

\subsubsection{Standard KV Cache}

\paragraph{Computational Complexity.} This method involves a one-time prefix computation and an iterative evaluation step over $E$ iterations.
\begin{itemize}
    \item \textbf{Prefix Computation (One-time):} A single forward pass over the prefix of length $N_p$ costs $\mathcal{O}(N_p^2)$.
    \item \textbf{Iterative Evaluation (per epoch):} For each of the $N_s+N_t$ tokens of the suffix and target, attention is computed over a context that grows from $N_p$ to $N_p+N_s+N_t-1$. The average context length for these tokens is approximately $N_p + (N_s+N_t)/2$. The computational cost for each of the $K \cdot q$ candidates is thus roughly proportional to the number of new tokens times twice the average context length, giving $\mathcal{O}((N_s + N_t) \cdot (2N_p + N_s + N_t))$.
\end{itemize}
The total complexity is the sum of the one-time cost and the total iterative cost:
\begin{align}
    C_{\text{Std}} &= \mathcal{O}(N_p^2) + E \cdot (K \cdot q) \cdot \mathcal{O}((N_s + N_t)(2N_p + N_s + N_t)) \nonumber \\
    &= \mathcal{O}(N_p^2 + E \cdot (K \cdot q) \cdot (N_s + N_t)(2N_p + N_s + N_t)).
\end{align}

\paragraph{Memory Complexity.} To achieve parallel computation, the pre-computed prefix cache of length $N_p$ must be physically duplicated $K \cdot q$ times. The total memory is the sum of these duplicated prefix caches and the caches for the suffix and target segments.
\begin{align}
    M_{\text{Std}} &= (K \cdot q) \cdot \mathcal{O}(N_p) + (K \cdot q) \cdot \mathcal{O}(N_s + N_t) \nonumber \\
    &= \mathcal{O}((K \cdot q) \cdot (N_p + N_s + N_t)).
\end{align}

\subsubsection{PSKV (Ours)}

\paragraph{Computational Complexity.} The computational process of PSKV is identical to the standard KV cache approach, as it also performs a one-time prefix computation followed by iterative evaluation of the suffix and target segments.
\begin{align}
    C_{\text{Ours}} = C_{\text{Std}} = \mathcal{O}(N_p^2 + E \cdot (K \cdot q) \cdot (N_s + N_t)(2N_p + N_s + N_t)).
\end{align}

\paragraph{Memory Complexity.} The key advantage of PSKV is that it avoids the physical duplication of the prefix cache. It stores the prefix cache of length $N_p$ only once, alongside the caches for the $K \cdot q$ suffix and target segments.
\begin{align}
    M_{\text{Ours}} &= \mathcal{O}(N_p) + (K \cdot q) \cdot \mathcal{O}(N_s + N_t) \nonumber \\
    &= \mathcal{O}(N_p + (K \cdot q) \cdot (N_s + N_t)).
\end{align}
This expression counts persistent KV-cache storage across layers. PSKV still uses a transient candidate-expanded prefix KV view inside each layer's attention computation, but this workspace is released after the layer is evaluated and is not stored across the full network stack.
\subsection{Derivations for Batched Complexity Analysis}
\label{appendix:complexity-batch}

In this section, we provide the derivations for the complexities of batched suffix synthesis presented in Section~\ref{sec:method:batch-suffix}. We let $B$ be the number of prompts, $E$ be the number of synthesis iterations, and $K \cdot q$ be the candidates per prompt per iteration. The total candidates per iteration is $B \cdot K \cdot q$. Let $N'_p, N_s, N'_t$ be the maximum token lengths of the aligned prefix, suffix, and target, and $N_{p, \text{avg}}, N_{t, \text{avg}}$ be the average lengths.

\subsubsection{Method without Acceleration}

\paragraph{Computational Complexity.} This method recomputes the full sequence for every candidate in every iteration. The total number of sequences processed is $E \cdot B \cdot K \cdot q$. The cost for each is quadratic to its length.
\begin{align}
    C_{\text{NoAccel}} &= E \cdot (B \cdot K \cdot q) \cdot \mathcal{O}((N_{p, \text{avg}} + N_s + N_{t, \text{avg}})^2) \nonumber \\
    &= \mathcal{O}(E \cdot (B \cdot K \cdot q) \cdot (N_{p, \text{avg}} + N_s + N_{t, \text{avg}})^2).
\end{align}

\paragraph{Memory Complexity.} The peak memory is determined by the batched inference step, which holds the KV caches for all $B \cdot K \cdot q$ candidates simultaneously.
\begin{align}
    M_{\text{NoAccel}} &= (B \cdot K \cdot q) \cdot \mathcal{O}(N_{p, \text{avg}} + N_s + N_{t, \text{avg}}) \nonumber \\
    &= \mathcal{O}((B \cdot K \cdot q) \cdot (N_{p, \text{avg}} + N_s + N_{t, \text{avg}})).
\end{align}

\subsubsection{Standard KV Cache}

\paragraph{Computational Complexity.} This method pre-computes the cache for the $B$ prefixes once, then iteratively evaluates the suffixes and targets.
\begin{itemize}
    \item \textbf{Prefix Computation (One-time):} A batched pass over $B$ prefixes of max length $N'_p$ costs $\mathcal{O}(B \cdot {N'_p}^2)$.
    \item \textbf{Iterative Evaluation (per epoch):} For each of the $B \cdot K \cdot q$ candidates, processing the $N_s+N'_t$ new tokens costs approximately $\mathcal{O}((N_s + N'_t) \cdot (2N'_p + N_s + N'_t))$.
\end{itemize}
The total complexity over $E$ iterations is the sum of the one-time cost and the total iterative cost:
\begin{align}
    C_{\text{Std}} = \mathcal{O}(B \cdot {N'_p}^2 + E \cdot (B \cdot K \cdot q) \cdot (N_s + N'_t)(2N'_p + N_s + N'_t)).
\end{align}

\paragraph{Memory Complexity.} To process in a single batch, the $B$ prefix caches must be duplicated $K \cdot q$ times each. The total memory is the sum of these duplicated prefix caches and the caches for the suffix/target segments.
\begin{align}
    M_{\text{Std}} &= (B \cdot K \cdot q) \cdot \mathcal{O}(N'_p) + (B \cdot K \cdot q) \cdot \mathcal{O}(N_s + N'_t) \nonumber \\
    &= \mathcal{O}((B \cdot K \cdot q) \cdot (N'_p + N_s + N'_t)).
\end{align}

\subsubsection{PSKV (Ours)}

\paragraph{Computational Complexity.} The computational process is identical to the standard KV cache approach.
\begin{align}
    C_{\text{Ours}} = C_{\text{Std}} = \mathcal{O}(B \cdot {N'_p}^2 + E \cdot (B \cdot K \cdot q) \cdot (N_s + N'_t)(2N'_p + N_s + N'_t)).
\end{align}

\paragraph{Memory Complexity.} PSKV avoids physical duplication. It stores one cache for each of the $B$ prefixes, and the caches for the $B \cdot K \cdot q$ suffix/target segments.
\begin{align}
    M_{\text{Ours}} &= B \cdot \mathcal{O}(N'_p) + (B \cdot K \cdot q) \cdot \mathcal{O}(N_s + N'_t) \nonumber \\
    &= \mathcal{O}(B \cdot N'_p + (B \cdot K \cdot q) \cdot (N_s + N'_t)).
\end{align}

\section{Computational Efficiency Analysis of AdvPrompter}
\label{app:advPrompter}

\subsection{Observations}

In our experimental evaluation of AdvPrompter (Table~\ref{tab:time-usage}), we observed two notable phenomena:

\begin{enumerate}
    \item \textbf{Severe performance degradation without KV caching:} All models exhibit substantially higher latency in the ``None'' (no acceleration) setting compared to cached configurations. For instance, on Llama-2-7B, runtime increases from 344 minutes with Standard KV Cache and 275 minutes with PSKV to 513 minutes in the None setting.
    
    \item \textbf{Model-specific latency disparities:} Among models of comparable size, Llama-2-7B and Llama-3-8B consistently exhibit higher execution times than Mistral-7B and Qwen2.5-7B across all acceleration settings. Even with our PSKV, Llama-2 (275 min) and Llama-3 (307 min) remain slower than Mistral (74 min) and Qwen (76 min).
\end{enumerate}

This section analyzes the algorithmic and architectural factors underlying these observations.

\subsection{The Cost of Cache-less Decoding}

The severe slowdown in the ``None'' setting stems from the fundamental inefficiency of autoregressive generation without KV caching.

\paragraph{Quadratic Total Complexity.}
With KV caching, the model avoids recomputing hidden states and K/V projections for previous tokens; each new token only attends to cached keys and values. Without caching, each decoding step recomputes the full prefix history, substantially increasing the total decoding cost.

\paragraph{Amplification by AdvPrompter's Search Strategy.}
Unlike gradient-based attacks (\emph{e.g.}, GCG) that perform single forward-backward passes, AdvPrompter employs beam search with sampling during suffix generation. This expands the effective batch size multiplicatively (batch $\times$ beam $\times$ samples), causing the quadratic recomputation penalty to be incurred across many parallel sequences simultaneously. The resulting computational burden saturates GPU throughput and explains the disproportionate slowdown observed in the ``None'' setting.

\subsection{Architectural Factors in Llama-Series Latency}

While all models benefit from KV caching, Llama-2 and Llama-3 exhibit consistently higher latency than Mistral and Qwen. We attribute this to two architecture-specific factors.

\paragraph{Llama-2: Multi-Head Attention Overhead.}
Llama-2-7B employs Multi-Head Attention (MHA) with 32 key-value heads, whereas Mistral-7B and Qwen2.5-7B utilize Grouped-Query Attention (GQA) with 8 key-value heads. In cache-less decoding, the model must recompute K and V projections for the entire sequence history at each generation step. With $4\times$ more KV heads, Llama-2 incurs proportionally higher memory bandwidth requirements for loading projection weights and storing intermediate KV tensors. This memory-bound overhead is particularly pronounced when batch sizes are large, as in AdvPrompter's beam search.

\paragraph{Llama-3: Vocabulary Projection Overhead.}
Llama-3-8B adopts GQA, mitigating the MHA-related overhead. However, it employs a vocabulary of 128K tokens---approximately $4\times$ larger than Llama-2 and Mistral ($\sim$32K). AdvPrompter uniquely requires training a generator model, which involves computing full logit distributions over the vocabulary and backpropagating through the output projection layer. The computational cost of this $\mathbb{R}^{d} \rightarrow \mathbb{R}^{|V|}$ projection and its gradient scales linearly with vocabulary size, imposing a persistent overhead on Llama-3 that cannot be alleviated by KV caching alone.

The ``None'' baseline forces redundant recomputation across the entire sequence history at each decoding step, a cost amplified by AdvPrompter's beam search strategy. Llama-2 is further penalized by its MHA architecture, which demands higher memory bandwidth than GQA-based models. Llama-3, while adopting GQA, suffers from the computational burden of its large vocabulary during AdvPrompter's generator training phase. Our PSKV effectively eliminates redundant prefix computations, providing consistent speedups across all evaluated models.

\section{Detailed analysis of memory overhead}
\label{app:memory_analysis}

To substantiate the memory efficiency analysis presented in the main text, we provide a detailed breakdown of the model architectures, the mathematical formulation of KV cache memory footprint, and the specific implementation differences between Standard KV Cache and PSKV.

\subsection{Architectural Specifications of Target Models}
The relative memory savings of PSKV depend on the attention mechanism and vocabulary size of the target Large Language Models (LLMs). Table~\ref{tab:model_specs} summarizes the critical architectural parameters for the models evaluated in this study.

\begin{table}[h]
\centering
\caption{Architectural specifications of the evaluated models. \textbf{MHA}: Multi-Head Attention; \textbf{GQA}: Grouped-Query Attention; \textbf{SWA}: Sliding Window Attention.}
\label{tab:model_specs}

{
\begin{tabular}{lccccc}
\toprule
\textbf{Model} & \textbf{Attention Type} & \textbf{Vocab Size} & \textbf{Q Heads} & \textbf{KV Heads} & \textbf{KV Compression} \\
\midrule
Vicuna-7B  & MHA & 32,000 & 32 & 32 & 1:1  \\
Llama-2-7B     & MHA & 32,000 & 32 & 32 & 1:1  \\
Llama-3-8B     & GQA & 128,256 & 32 & 8  & 4:1 \\
Mistral-7B& GQA + SWA & 32,000 & 32 & 8  & 4:1 \\
Qwen2.5-7B     & GQA & 151,936 & 28 & 4  & 7:1 \\
\bottomrule
\end{tabular}
}
\end{table}

\paragraph{\textbf{Impact of Attention Mechanisms (MHA vs. GQA).}}
The memory footprint of the KV cache is determined by the number of Key-Value heads.
\begin{itemize}
\item \textbf{Multi-Head Attention (MHA):} Used by Llama-2 and Vicuna. Here, $H_{kv}=H_q$. Every query head requires a unique key-value pair, leading to maximum memory usage.
\item \textbf{Grouped-Query Attention (GQA):} Used by Llama-3, Mistral, and Qwen. Here, $H_{kv}<H_q$. Multiple query heads share a single KV head (e.g., Qwen utilizes 28 query heads but only 4 KV heads). This architectural change physically reduces the KV cache tensor size by a factor of $H_q/H_{kv}$, which inherently lowers the memory ceiling and thus reduces the \textit{relative} impact of PSKV compared to MHA models.
\end{itemize}

\paragraph{\textbf{Impact of Vocabulary Size.}}
The memory consumption during attack generation is not solely due to the KV cache but also intermediate activation tensors, specifically the logits tensor, which has shape $B \times T \times |\mathcal{V}|$. As shown in Table~\ref{tab:model_specs}, Llama-3 and Qwen2.5 have vocabulary sizes $|\mathcal{V}|$ approximately $4\times$ to $5\times$ larger than Llama-2. This drastically increases the non-KV memory overhead (e.g., storing logits and gradients), further diluting the proportion of memory saved by PSKV.

\paragraph{\textbf{Note on Mistral's Sliding Window.}}
Mistral-7B employs Sliding Window Attention (SWA) with a large window size. While SWA theoretically caps the KV cache size to the window length, our experiments use suffix attack sequences whose total length is smaller than this window. Therefore, the cache behavior in our experiments effectively mirrors standard GQA without the sliding window truncation benefits.

\subsection{Theoretical Memory Complexity: Standard vs. PSKV}

We formalize the memory consumption to explain why Standard KV Cache results in OOM errors under beam search settings while PSKV remains efficient.
Let $B$ be the batch size (or beam width/number of candidates), $L_{pre}$ be the prefix length, $L_{gen}$ be the generated-token length, $D$ be the hidden dimension, $H_{kv}$ be the number of KV heads, and $N_{layer}$ be the number of layers.

\paragraph{\textbf{Standard KV Cache (Eager Expansion).}}
Standard implementations (e.g., Hugging Face default) eagerly expand the KV cache to match the batch dimension at initialization. The memory complexity for the cached prefix is:
\begin{equation}
M_{std} \propto N_{layer} \times \mathbf{B} \times L_{pre} \times H_{kv} \times D
\end{equation}
Under optimization-based attacks (e.g., GCQ, BEAST), $B$ represents the number of candidates. The $B \times L_{pre}$ term causes memory usage to scale linearly with the candidate count, leading to rapid exhaustion of GPU memory.

\paragraph{\textbf{PSKV (Layer-wise Dynamic Expansion).}}
Our PSKV implementation maintains a singleton copy of the prefix cache. During the forward pass of the $\ell$-th layer, the cache is virtually broadcasted or temporarily expanded only for that specific layer's computation and immediately released. The resident memory footprint is:
\begin{equation}
M_{pskv} \propto N_{layer} \times \mathbf{1} \times L_{pre} \times H_{kv} \times D
\end{equation}
The temporary expansion memory required during the forward pass is $\mathcal{O}(B \cdot L_{pre} \cdot H_{kv} \cdot D)$ for the current layer computation. This allows the resident memory consumption related to the prefix to remain independent of the attack batch size $B$, providing the efficiency gains observed in the main results.

\subsection{Attack-Specific Memory Overhead Details}
Different attack algorithms impose different auxiliary memory costs that interact with the cache mechanism:
\begin{itemize}
\item \textbf{High-Width Search (GCQ, BEAST):} These algorithms generate a large number of candidate sequences per step. While PSKV compresses the prefix storage, the intermediate activations (e.g., attention scores, FFN intermediate states) still scale with $B$. Thus, while PSKV prevents OOM from the \textit{cache} perspective, the \textit{activation} memory creates a new floor for memory usage.
\item \textbf{Dual-Model Loading (AdvPrompter):} AdvPrompter requires the simultaneous residence of the Target LLM and the Attacker LLM (Prompter) in VRAM. This effectively halves the available memory for dynamic tensors. In this constrained environment, the static weight of the models dominates the memory profile, making the dynamic savings from PSKV less visually prominent in percentage terms compared to single-model attacks.
\end{itemize}

\section{ASR Stability and Numerical Equivalence}
\label{app:asr-stability}

PSKV is an efficiency-only modification to the inference procedure. It reuses the KV tensors of the shared prefix to reduce redundant computation, while leaving the attack objective, candidate generation rule, search strategy, decoding policy, and evaluation metric unchanged. Thus, for the same model inputs and candidate suffixes, PSKV is expected to preserve the semantics of the original attack pipeline.

In the main experiments, ASR is reported as the average over three independent runs. This reduces the influence of run-to-run randomness, which is common in suffix-based attacks due to random initialization, candidate sampling, tie-breaking, and stochastic generation or optimization steps. Small ASR differences between execution modes should therefore be interpreted relative to this natural variance.

We further conduct two checks to verify that the observed ASR variation is not caused by an approximation introduced by PSKV. First, we run 10 additional GCG experiments on the same data, paired by random seed, and compare PSKV against No Cache. A paired $t$-test gives $t = 0.35$ and $p = 0.74$, with a 95\% confidence interval of $[-0.022, +0.030]$ and Cohen's $d = 0.12$. These statistics indicate that the ASR difference between PSKV and No Cache is not statistically significant and has a negligible effect size. Therefore, the observed fluctuations are consistent with the inherent stochasticity of GCG candidate sampling rather than a systematic change introduced by PSKV.

Second, we compare the output logits of No Cache, Standard KV Cache, and PSKV on identical inputs. Under FP32/FP64 execution, all three modes produce bit-identical logits. Under BF16 execution, minor rounding discrepancies can appear between cached and uncached inference due to reduced-precision arithmetic. However, Standard KV Cache and PSKV always produce exactly identical logits under the same BF16 setting. This confirms that PSKV is numerically equivalent to standard KV-cache execution. In other words, the optimization landscape observed by GCG is unchanged when replacing Standard KV Cache with PSKV, and any ASR variation originates from the stochastic components of the attack itself. We use BF16 in the main experiments because FP32 increases wall-clock time by approximately $8\times$ while producing no measurable impact on ASR.

Overall, PSKV should be interpreted as an inference acceleration layer rather than a modification to the attack algorithm. It changes the execution of shared-prefix computation, but not the algorithmic behavior of the evaluated attacks.

\section{Scaling to Longer Suffixes and Longer Prompts}
\label{app:longer-contexts}

We further evaluate how PSKV scales under longer candidate suffixes and longer instruction prompts. These settings stress different parts of the computation. Increasing the suffix length mainly increases candidate-specific forward and gradient computation, whereas increasing the prefix length directly amplifies prefix processing and prefix-cache storage. In addition to peak-memory and runtime measurements, we provide CUDA memory snapshots to visualize the allocation patterns of different execution strategies. In these snapshots, the vertical scale represents allocated GPU memory, and persistent colored blocks indicate tensors that remain resident over long intervals, while narrow spikes indicate transient allocations. The snapshots are used to illustrate memory behavior qualitatively; the peak-memory numbers are reported in Table~\ref{tab:longer_contexts}.

\begin{figure*}[htbp]
\centering
\textbf{(a) No Cache, SFX20}

\vspace{4pt}
\includegraphics[width=0.95\textwidth]{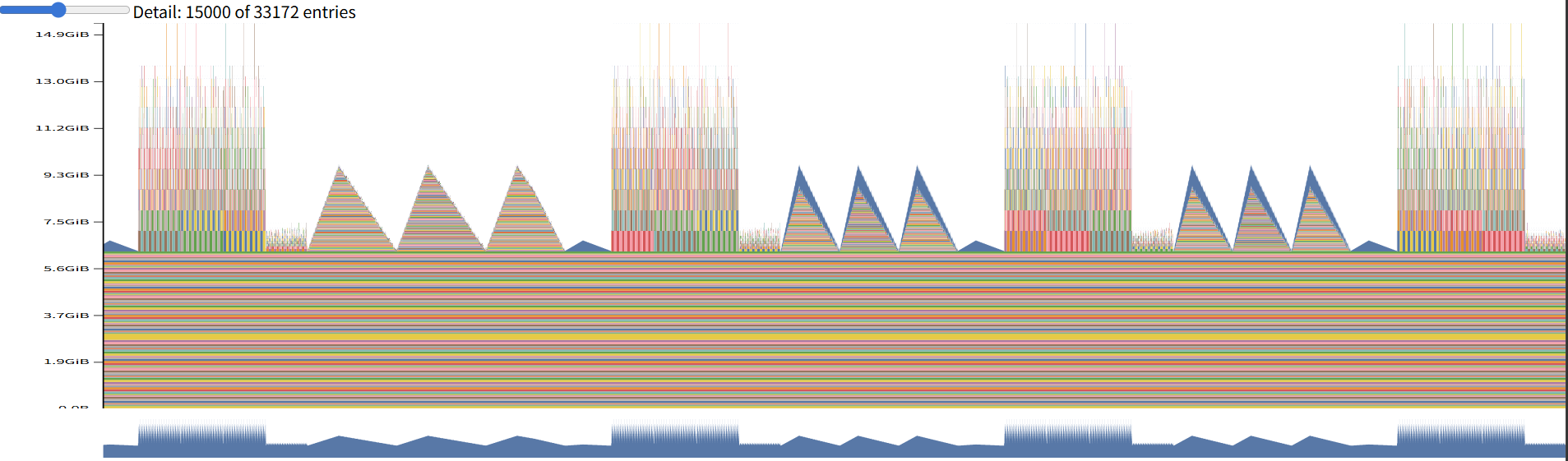}

\vspace{8pt}
\textbf{(b) Standard KV Cache, SFX20}

\vspace{4pt}
\includegraphics[width=0.95\textwidth]{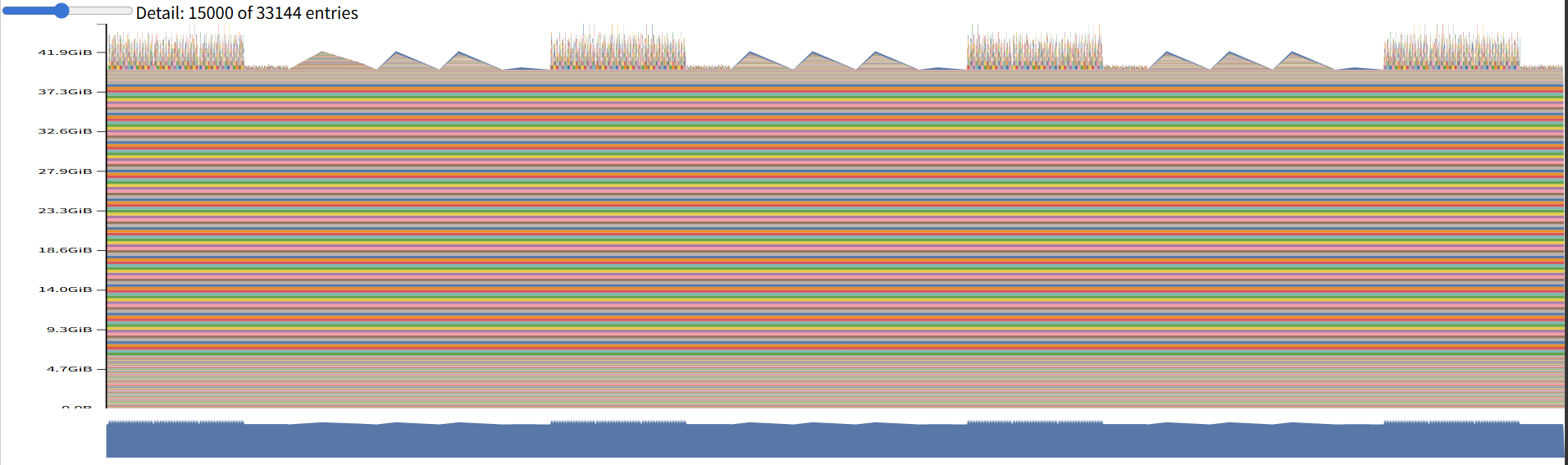}

\vspace{8pt}
\textbf{(c) PSKV, SFX20}

\vspace{4pt}
\includegraphics[width=0.95\textwidth]{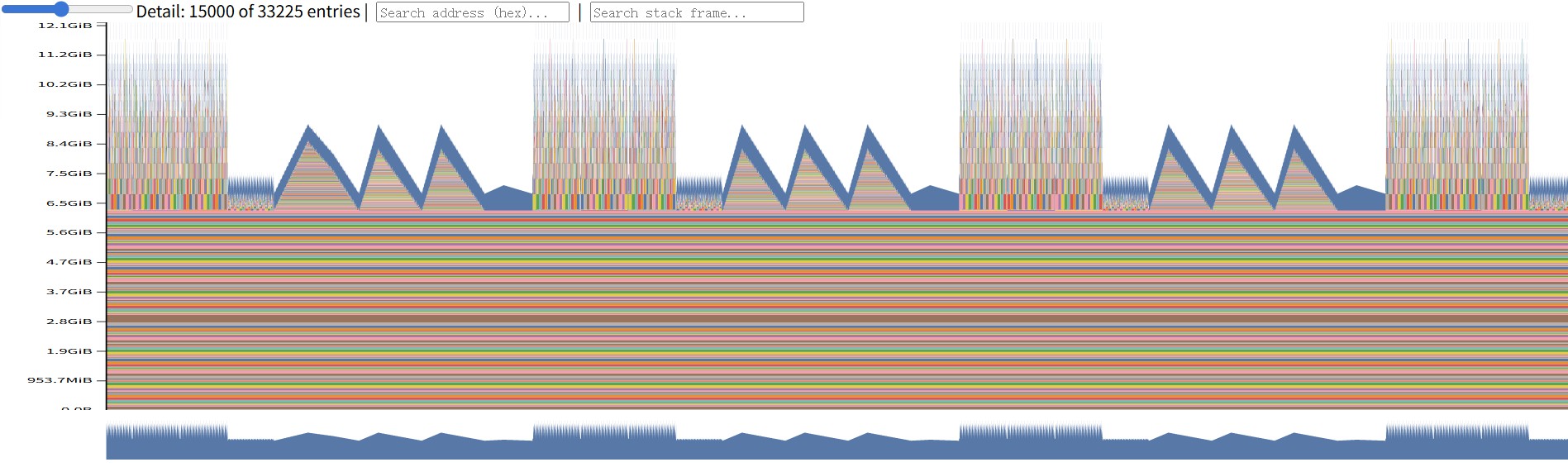}

\caption{Regular-setting memory snapshots under SFX20. No Cache exhibits transient peaks during full-sequence gradient forward passes. Standard KV Cache keeps a candidate-expanded prefix KV cache resident, forming a persistent high-memory plateau. PSKV avoids both behaviors by storing one compact prefix cache and using only transient layer-wise KV expansion.}
\label{fig:memory_snapshot_regular}
\end{figure*}

\paragraph{Regular memory behavior.}
We first examine the regular SFX20 setting. Figure~\ref{fig:memory_snapshot_regular} shows that the three strategies have qualitatively different memory traces. No Cache does not maintain a prefix cache, but it repeatedly performs full-sequence forward and gradient computation for every candidate. As a result, its memory trace contains large transient peaks during gradient forward passes, where the prefix, suffix, and target tensors are materialized together for the candidate batch.

Standard KV Cache reduces redundant prefix computation, but it keeps a fully expanded prefix KV cache across the candidate batch. This creates a persistent high-memory plateau that remains resident throughout candidate evaluation. In contrast, PSKV avoids both patterns. It stores only one compact prefix KV cache and materializes the candidate-expanded prefix KV view only inside the current layer's attention computation. The expanded view is discarded immediately after the layer is evaluated. Therefore, compared with No Cache, PSKV reduces the large forward-pass memory peaks; compared with Standard KV Cache, it avoids the persistent resident memory occupied by expanded cache tensors.

\paragraph{Scaling to longer suffixes.}
Table~\ref{tab:longer_contexts} reports peak memory usage and runtime of GCG on Llama-2-7B under SFX40 and SFX60. As the suffix length increases, all strategies become slower because longer suffixes require longer candidate-specific forward and gradient passes. Memory usage also increases because suffix-side activations, attention workspaces, and loss-computation tensors become larger.

\begin{figure*}[htbp]
\centering
\textbf{(a) No Cache, SFX40}

\vspace{4pt}
\includegraphics[width=0.95\textwidth]{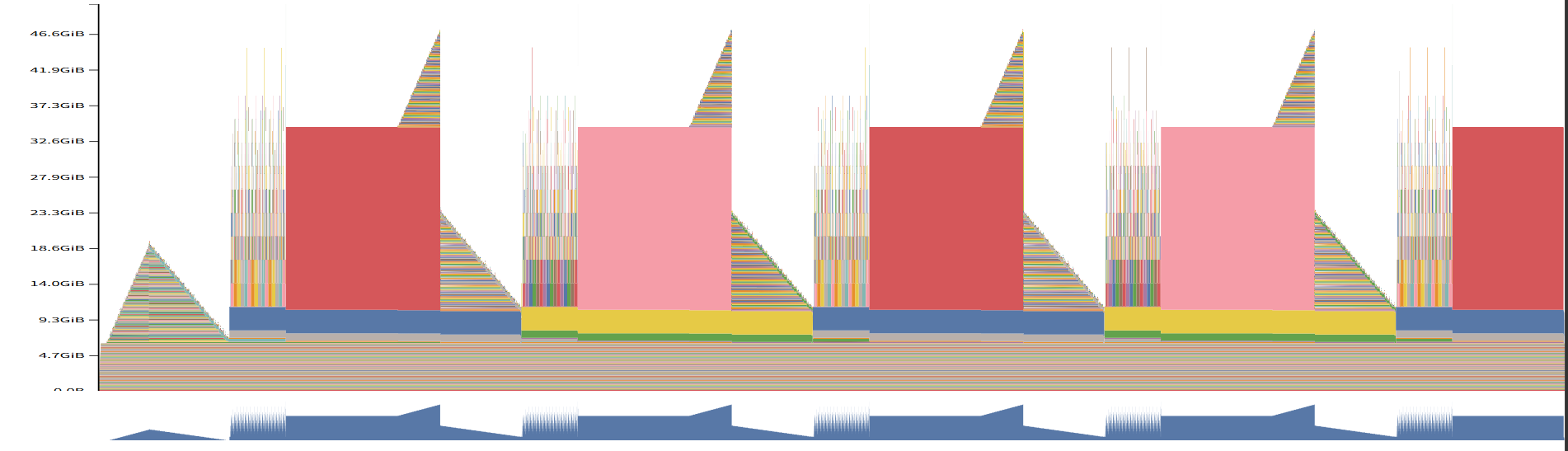}

\vspace{8pt}
\textbf{(b) Standard KV Cache, SFX40}

\vspace{4pt}
\includegraphics[width=0.95\textwidth]{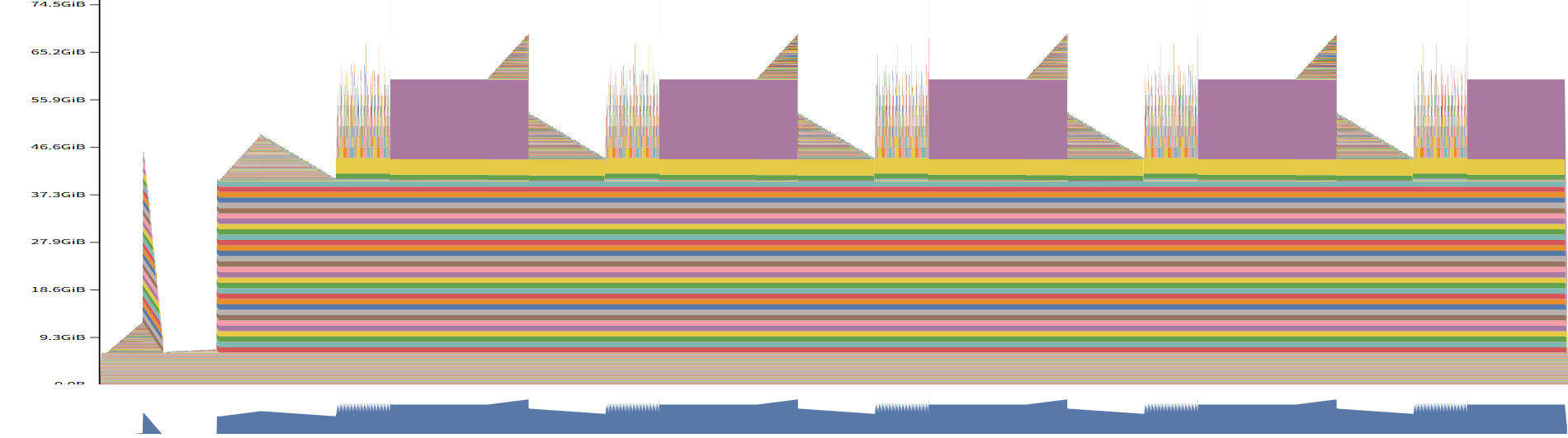}

\vspace{8pt}
\textbf{(c) PSKV, SFX40}

\vspace{4pt}
\includegraphics[width=0.95\textwidth]{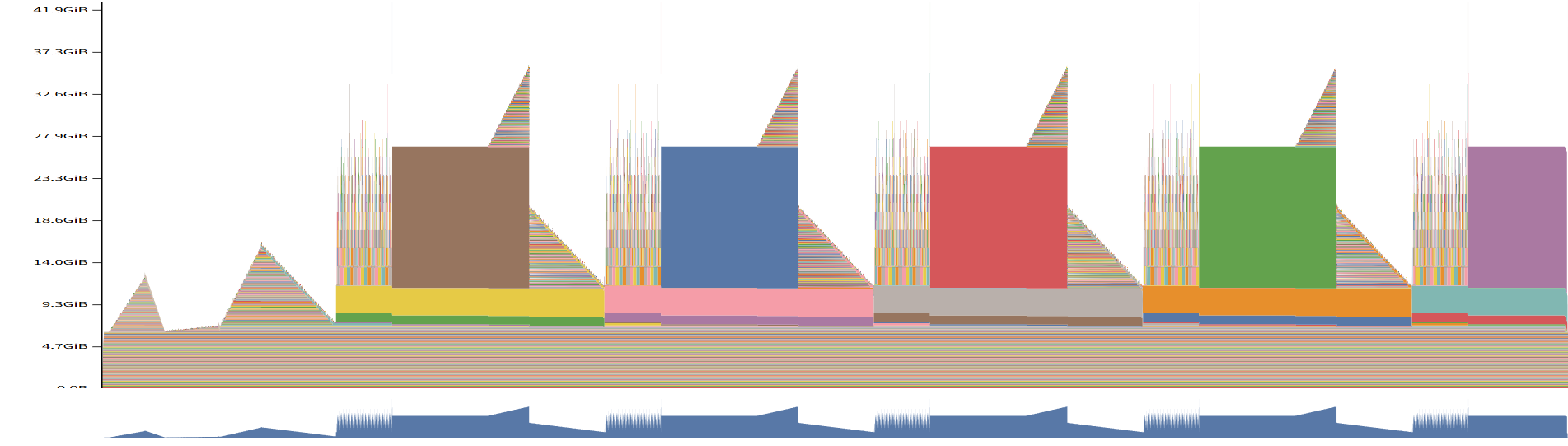}

\caption{Memory snapshots under SFX40. No Cache produces large transient peaks during full-sequence gradient forward passes, while Standard KV Cache forms a persistent high-memory plateau by maintaining a fully expanded prefix KV state. PSKV avoids both behaviors by keeping the prefix cache compact and expanding it only as transient per-layer workspace.}
\label{fig:memory_snapshot_sfx40}
\end{figure*}

The memory snapshots in Figure~\ref{fig:memory_snapshot_sfx40} further show the difference between the three strategies under SFX40. No Cache produces repeated transient peaks from full-sequence gradient forward passes. Standard KV Cache forms a persistent high-memory plateau because the expanded prefix KV cache remains resident across the candidate batch. PSKV avoids this persistent cache plateau and also reduces the forward-pass peaks relative to No Cache, since the shared prefix is cached compactly and expanded only layer by layer.

\begin{figure*}[htbp]
\centering
\textbf{(a) No Cache, SFX60}

\vspace{4pt}
\includegraphics[width=0.95\textwidth]{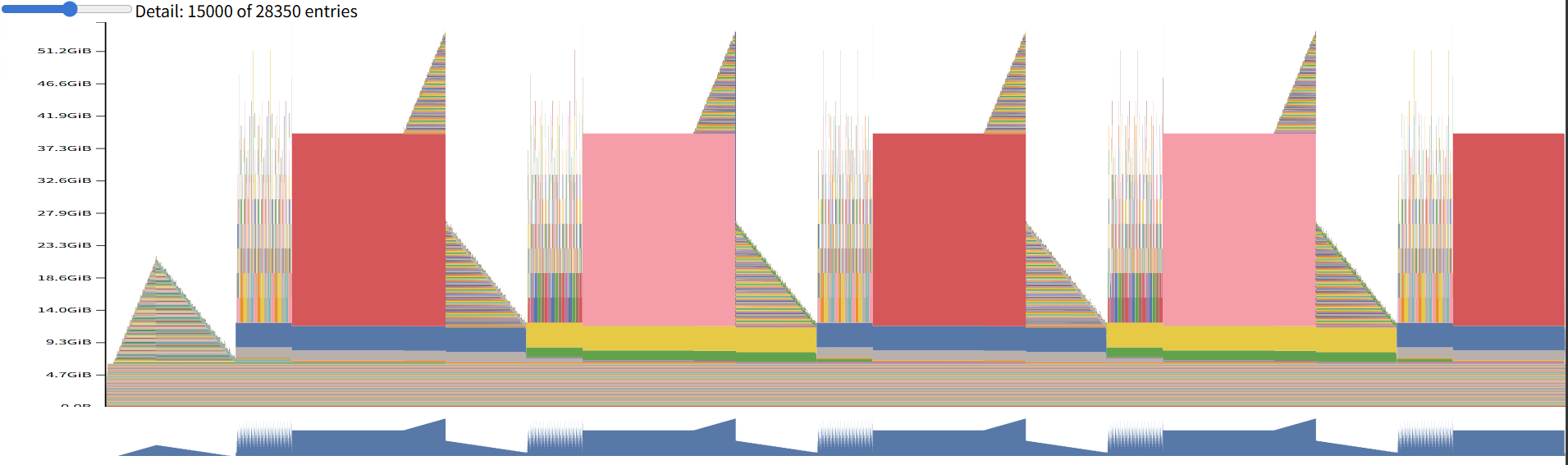}

\vspace{8pt}
\textbf{(b) PSKV, SFX60}

\vspace{4pt}
\includegraphics[width=0.95\textwidth]{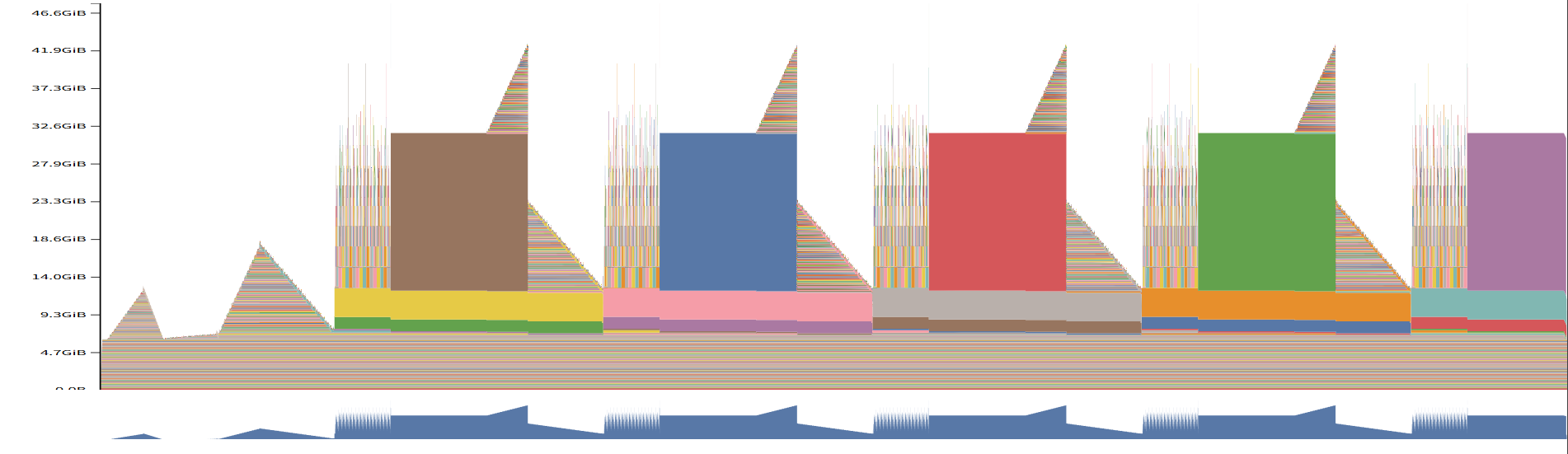}

\caption{Memory snapshots under SFX60. Longer suffixes increase candidate-specific forward and gradient cost. No Cache shows larger full-forward memory peaks, whereas PSKV continues to avoid persistent candidate-expanded prefix KV storage through layer-wise lazy expansion. The Standard KV Cache snapshot is omitted because the additional profiling overhead causes the fully expanded cache execution to run out of memory.}
\label{fig:memory_snapshot_sfx60}
\end{figure*}

Under SFX60, the suffix-side computation becomes even more expensive. Figure~\ref{fig:memory_snapshot_sfx60} compares No Cache and PSKV. The Standard KV Cache snapshot is omitted because the additional memory overhead from CUDA memory snapshot profiling causes the fully expanded prefix-cache execution to run out of memory in this setting. The non-profiled Standard KV run already reaches a high aggregate peak memory, as shown in Table~\ref{tab:longer_contexts}. This behavior is consistent with the SFX40 snapshot: Standard KV Cache is sensitive to candidate-expanded prefix storage, while PSKV keeps the expanded prefix KV tensors transient.

\begin{table}[htbp]
\centering
\caption{GCG scaling with longer suffixes and prompts on Llama-2-7B. Entries are memory in GB / time in minutes.}
\label{tab:longer_contexts}
\begin{tabular}{lccc}
\toprule
\textbf{Setting} & \textbf{No Cache} & \textbf{Standard KV} & \textbf{PSKV} \\
\midrule
SFX40 & 113 / 255 & 140 / 190 & 80 / 190 \\
SFX60 & 116 / 306 & 147 / 230 & 94 / 230 \\
SFX20--WildJailbreak & 126 / 363 & OOM & 64 / 206 \\
\bottomrule
\end{tabular}
\end{table}

\paragraph{Scaling to longer prompts.}
We also evaluate longer-prefix inputs from WildJailbreak~\citep{jiang2024wildteaming}. This setting stresses prefix-side memory more directly than increasing suffix length. For No Cache, the full prompt is repeatedly processed for every candidate, so the memory peak during the gradient forward pass grows with prefix length. For Standard KV Cache, the prefix computation is reused, but the full prefix KV cache must be expanded across the candidate batch and kept resident across layers. This makes Standard KV Cache especially sensitive to prefix length increases. As shown in Table~\ref{tab:longer_contexts}, Standard KV Cache runs out of memory in the SFX20--WildJailbreak setting, and therefore no complete Standard KV memory snapshot is available.

\begin{figure*}[htbp]
\centering
\textbf{(a) No Cache, SFX20--WildJailbreak}

\vspace{4pt}
\includegraphics[width=0.95\textwidth]{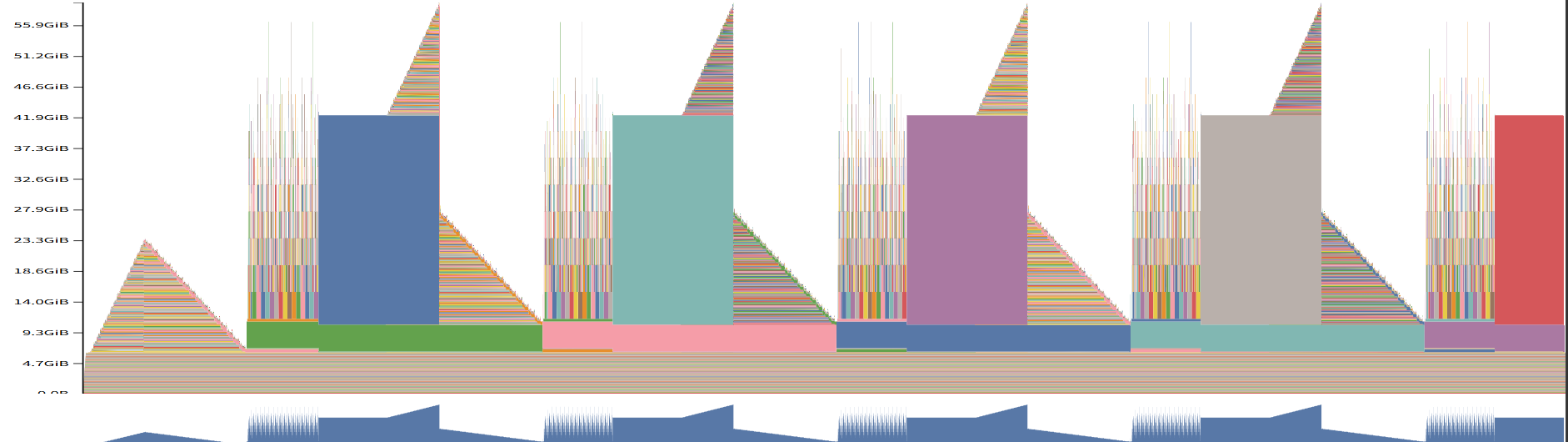}

\vspace{8pt}
\textbf{(b) PSKV, SFX20--WildJailbreak}

\vspace{4pt}
\includegraphics[width=0.95\textwidth]{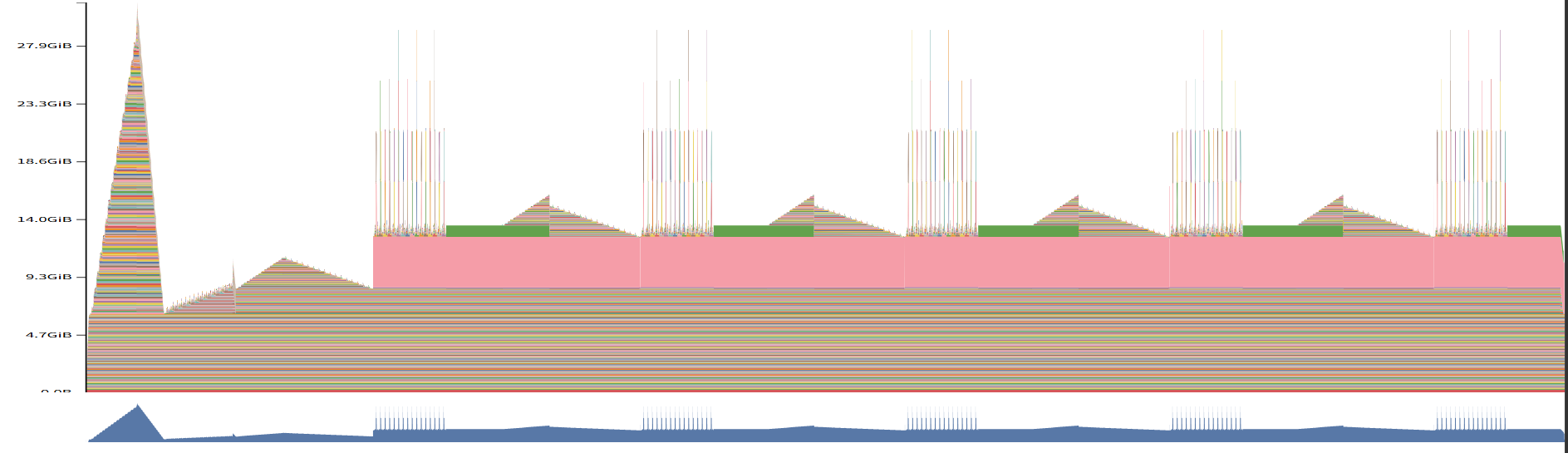}

\caption{Memory snapshots under longer-prefix WildJailbreak prompts. No Cache exhibits large gradient-forward peaks because the long prefix is repeatedly processed for every candidate. Standard KV Cache runs out of memory because it must keep a fully expanded long-prefix KV cache resident across the candidate batch. PSKV keeps only one compact prefix KV cache and uses transient layer-wise expansion, allowing it to remain memory-efficient in this long-prefix setting.}
\label{fig:memory_snapshot_wildjailbreak}
\end{figure*}

PSKV avoids this persistent expanded-prefix state. It stores only one compact prefix KV cache per prompt and materializes the candidate-expanded KV view only during the current layer's attention computation. Consequently, PSKV achieves lower peak memory than No Cache and avoids the memory explosion of Standard KV Cache. Figure~\ref{fig:memory_snapshot_wildjailbreak} shows that No Cache suffers from large long-prefix forward peaks, while PSKV maintains a much smaller resident footprint with only transient layer-wise expansion. This indicates that the advantage of PSKV becomes more pronounced as the shared prefix becomes longer.

Overall, these results and memory snapshots support the main memory-efficiency claim in Section~\ref{sec:experiments}. No Cache suffers from large memory peaks during full-sequence gradient forward passes, while Standard KV Cache suffers from persistent memory growth by maintaining fully expanded prefix KV states. PSKV addresses both issues: it reuses the prefix computation through a compact shared KV cache and keeps candidate-expanded KV tensors transient through layer-wise lazy expansion. This keeps the resident footprint minimal and makes PSKV increasingly advantageous under larger suffix-search workloads and longer-prefix prompts. 

\section{Using and Implementing the PSKV Framework}
\label{app:using-pskv}

This appendix describes how to integrate PSKV into existing suffix-based attack pipelines. PSKV is implemented as a modular inference acceleration framework. It does not require changes to the attack objective, candidate generation procedure, search strategy, or loss function. Instead, it replaces repeated full-sequence forward passes with prefix-aware cached forward passes.

\subsection{Integration Overview}

Many suffix-based attack algorithms repeatedly evaluate multiple candidate suffixes under the same instruction prefix. Given an input sequence consisting of a shared prefix and a candidate-specific suffix, the standard implementation usually performs a full forward pass for every candidate. PSKV avoids redundant prefix computation by first computing the KV cache of the shared prefix and then reusing this cache when evaluating different suffix candidates.

To use PSKV, the attack pipeline only needs to separate the input into two parts:
\begin{itemize}
    \item a shared prefix, which remains unchanged across candidate evaluations;
    \item one or more candidate suffixes, which are evaluated under the shared prefix.
\end{itemize}

After this separation, the original forward computation can be replaced by the PSKV cached forward interface.

\subsection{Basic Usage}

The typical integration consists of three steps.

\paragraph{Step 1: Initialize the prefix cache.}
Before evaluating candidate suffixes, PSKV first computes and stores the KV cache for the shared prefix:

\begin{verbatim}
cache = initialize_prefix_cache(
    model=model,
    prefix_input_ids=prefix_input_ids,
    prefix_attention_mask=prefix_attention_mask,
    cache_mode="pskv",
)
\end{verbatim}

The prefix cache only needs to be recomputed when the shared prefix changes. If multiple candidate suffixes are evaluated for the same instruction prefix, the same cache can be reused across candidate-scoring calls.

\paragraph{Step 2: Replace the standard forward call.}
A standard implementation usually evaluates each full input sequence with a regular model forward pass:

\begin{verbatim}
outputs = model(
    input_ids=full_input_ids,
    attention_mask=full_attention_mask,
)
logits = outputs["logits"]
\end{verbatim}

With PSKV, this call is replaced by a cached forward pass over the suffix-side inputs:

\begin{verbatim}
outputs = forward_with_cache(
    model=model,
    cache=cache,
    suffix_input_ids=suffix_input_ids,
    suffix_attention_mask=suffix_attention_mask,
    cache_mode="pskv",
)
logits = outputs["logits"]
\end{verbatim}

The returned logits can then be passed to the original loss or scoring function without changing the attack-specific optimization logic.

\paragraph{Step 3: Set the cache expansion factor.}
When multiple suffix candidates are evaluated in parallel, PSKV sets an expansion factor so that each layer creates a transient candidate-batched view of the compact prefix cache during attention computation:

\begin{verbatim}
outputs = forward_with_cache(
    model=model,
    cache=cache,
    suffix_input_ids=suffix_input_ids,
    suffix_attention_mask=suffix_attention_mask,
    cache_mode="pskv",
    expand_factor=num_candidates,
)
\end{verbatim}

Here, \texttt{num\_candidates} denotes the number of suffix candidates scored in the current batch. This allows PSKV to reuse one shared-prefix cache while evaluating multiple candidate suffixes simultaneously.

\subsection{Example: Integrating PSKV into GCG}

Algorithm~\ref{alg:pskv-gcg-integration} illustrates the integration pattern using GCG as an example. The original GCG procedure is kept unchanged, including candidate sampling, candidate scoring, and suffix update. PSKV is only used to accelerate the forward computation during candidate scoring.

\begin{algorithm}[htb]
\caption{Integrating PSKV into a suffix-based attack pipeline}
\label{alg:pskv-gcg-integration}
\begin{algorithmic}[1]
\Require Model $f_\theta$, shared prefix $x_{\mathrm{pre}}$, initial suffix $x_{\mathrm{suf}}$, number of candidates $K$
\State $C \gets \Call{InitializePrefixCache}{f_\theta, x_{\mathrm{pre}}}$
\While{attack not converged}
    \State $\mathcal{S} \gets \Call{GenerateCandidates}{x_{\mathrm{suf}}, K}$
    \State Expand $C$ to match the candidate batch size $K$
    \State $Y \gets \Call{ForwardWithCache}{f_\theta, C, \mathcal{S}}$
    \State $\ell \gets \Call{ComputeAttackLoss}{Y}$
    \State $x_{\mathrm{suf}} \gets \Call{UpdateSuffix}{\mathcal{S}, \ell}$
\EndWhile
\State \Return $x_{\mathrm{suf}}$
\end{algorithmic}
\end{algorithm}

\subsection{Position Consistency under Padding and RoPE}
\label{app:position-consistency}

PSKV changes only the tensor layout used for batching and cache reuse; it does not change the logical token positions seen by the model. In our preprocessing, the instruction prefix is left-padded, while the suffix-side continuation and target sequence are left-aligned. The prefix mask records the real prefix tokens, and padding tokens are excluded from attention. In the implementation, the prefix tensor is filled from the right and the corresponding \texttt{message\_mask} is set to one on real prefix tokens, while the target-side tensor is filled from the left and \texttt{target\_mask} is set to one only on true target-response tokens.

For sample $i$, let $m_i \in \{0,1\}^{N'_p}$ be the prefix attention mask. The logical position of a real prefix token at padded index $j$ is
\begin{align}
p^{\mathrm{pre}}_{i,j}
=
\sum_{u\leq j} m_{i,u} - 1,
\qquad \text{for } m_{i,j}=1.
\end{align}
Let $n_i=\sum_j m_{i,j}$ be the true prefix length. Then the local suffix/target-side token at index $r$ receives position
\begin{align}
p^{\mathrm{suf/tar}}_{i,r}=n_i+r.
\end{align}
These are exactly the positions that the same tokens would have in the unpadded sequence
$x_i^{(h)} \oplus x_i^{(s)} \oplus y_i^{(h)}$.
For RoPE-based models, this preserves the same relative phase relationships as the full-sequence execution. PSKV relies on the model's standard \texttt{attention\_mask}, cache length, and \texttt{past\_key\_values} logic to construct the same position offsets during cached execution.

The corresponding preprocessing can be summarized as:
\begin{verbatim}
# Prefix side: left padding.
message_ids[i, -len(prefix_i):] = prefix_i
message_mask[i, -len(prefix_i):] = 1

# Suffix/target side: left-aligned.
target_ids[i, :len(after_i) + len(target_i)] = concat(after_i, target_i)

# Loss mask: only target-response tokens contribute to loss.
target_mask[i, len(after_i):len(after_i) + len(target_i)] = 1
\end{verbatim}

\subsection{Layer-Wise Lazy Expansion and Its Overhead}
\label{app:lazy-expansion-overhead}

PSKV implements lazy expansion inside each cache layer. The compact prefix cache is stored once as \texttt{self.keys} and \texttt{self.values}. During candidate scoring, PSKV sets an expansion factor through the cached forward interface. Each layer then creates a broadcasted view of the compact prefix cache using \texttt{torch.expand}, reshapes it to the candidate batch dimension, and concatenates it with the suffix/target K/V tensors for the current layer. The concatenated tensor is used only as the current layer's attention workspace and is not written back to the persistent cache.

This differs from Standard KV Cache. Standard KV Cache expands the prefix cache to the full search width during initialization and materializes a contiguous candidate-expanded tensor for every layer. In contrast, PSKV stores only the compact prefix cache persistently and materializes candidate-expanded K/V tensors only as transient per-layer workspaces.

Let $L$ be the number of layers, $N_{\mathrm{cand}}$ the candidate batch size, $B$ the prompt batch size, $N_p$ the prefix length, $L_{\mathrm{dec}}$ the suffix/target length, $H_{kv}$ the number of KV heads, $d_h$ the head dimension, and $b$ the number of bytes per element. Standard KV Cache stores a persistent expanded prefix cache of size
\begin{align}
M_{\mathrm{Std,prefix}}
=
\mathcal{O}\!\left(
L\,N_{\mathrm{cand}}\,N_p\,H_{kv}\,d_h\,b
\right).
\end{align}
PSKV instead stores the compact prefix cache persistently:
\begin{align}
M_{\mathrm{PSKV,prefix}}
=
\mathcal{O}\!\left(
L\,B\,N_p\,H_{kv}\,d_h\,b
\right),
\end{align}
and uses a transient per-layer workspace of size
\begin{align}
M_{\mathrm{temp/layer}}
=
\mathcal{O}\!\left(
N_{\mathrm{cand}}\,(N_p+L_{\mathrm{dec}})\,H_{kv}\,d_h\,b
\right).
\end{align}
The additional lazy-expansion copy/bandwidth cost is dominated by constructing the per-layer concatenated K/V tensors:
\begin{align}
T_{\mathrm{lazy}}
=
\mathcal{O}\!\left(
L\,N_{\mathrm{cand}}\,N_p\,H_{kv}\,d_h
\right).
\end{align}
This cost is a memory-bandwidth overhead, not an additional attention computation term. In our workloads, it is small relative to the candidate scoring, attention, and loss-computation cost, which is consistent with the similar runtime of PSKV and Standard KV Cache in Table~\ref{tab:time-usage}.

The memory snapshots in Appendix~\ref{app:longer-contexts} empirically confirm this implementation behavior. In the regular SFX20 setting, No Cache exhibits transient peaks during full-sequence gradient forward passes, because the prefix, suffix, and target tensors are repeatedly materialized for each candidate batch. Standard KV Cache shows a persistent high-memory plateau, corresponding to the candidate-expanded prefix KV cache that remains resident across layers. In contrast, PSKV avoids both behaviors: it stores only one compact prefix cache persistently and uses only transient layer-wise KV expansion during attention computation. This snapshot pattern is consistent with the implementation, where \texttt{torch.expand} creates a broadcasted view and only the current layer's concatenated K/V workspace is materialized and released after use.

\subsection{Interaction with Multi-GPU Parallelism}
\label{app:multi-gpu-cache}

PSKV does not introduce a separate distributed cache manager. It initializes prefix caches from the model's own \texttt{past\_key\_values}; therefore, the cache placement follows the device placement of the underlying model. Under layer-wise model or pipeline parallelism, each GPU stores only the prefix K/V tensors for its local layers. Under tensor parallelism, each GPU stores only its local KV-head or hidden-dimension shard. Under data parallelism, each rank stores the prefix caches for its local prompt batch. PSKV does not perform additional all-gather or cross-GPU synchronization of prefix caches.

Thus, the memory reduction applies locally on each GPU. For GPU $g$, let $\mathcal{L}_g$ be the layers placed on that GPU, $H_{kv}^{(g)}$ the local number of KV heads, and $B_g$ the local prompt batch size. The persistent prefix-cache memory of PSKV scales as
\begin{align}
M_{\mathrm{PSKV}}^{(g)}
=
\mathcal{O}\!\left(
\sum_{\ell\in\mathcal{L}_g}
B_g\,N_p\,H_{kv}^{(g)}\,d_h\,b
\right)
+
\text{transient per-layer workspace}.
\end{align}
By contrast, Standard KV Cache scales with the local candidate batch size:
\begin{align}
M_{\mathrm{Std}}^{(g)}
=
\mathcal{O}\!\left(
\sum_{\ell\in\mathcal{L}_g}
N_{\mathrm{cand},g}\,N_p\,H_{kv}^{(g)}\,d_h\,b
\right).
\end{align}

\subsection{Ragged Targets, Loss Masking, and Gradient Flow}
\label{app:ragged-target-loss}

For batched multi-instruction evaluation, target responses can have different lengths. PSKV handles this with an explicit target mask. For each sample $i$, let $a_i$ denote the post-suffix template continuation and $y_i$ the target response. We construct
\begin{align}
z_i = a_i \oplus y_i
\end{align}
and left-align $z_i$ in the suffix-side tensor. The loss mask is set to one only on the target-response span:
\begin{align}
m^{\mathrm{tar}}_{i,r}
=
\begin{cases}
1, & |a_i| \le r < |a_i| + |y_i|,\\
0, & \text{otherwise}.
\end{cases}
\end{align}
The objective is then computed as a masked average:
\begin{align}
\mathcal{L}
=
\frac{
\sum_{i,c,r}
m^{\mathrm{tar}}_{i,r}\,
\mathrm{CE}(\ell_{i,c,r-1}, z_{i,r})
}{
\sum_{i,c,r} m^{\mathrm{tar}}_{i,r}
}.
\end{align}
Here $c$ indexes candidate suffixes. Template continuation tokens and padding tokens are included in the context when needed, but they receive zero loss weight. Therefore, gradients flow only from valid target-token losses, while suffix-side tensors remain dense and vectorized across candidates and instructions.

\section{Limitations}
\label{app:limitation}
PSKV targets workloads with a shared-prefix, many-suffix structure. Its gains are smaller when prefix lengths are short, candidate widths are low, or non-cache tensors such as logits, activations, or dual-model training state dominate memory. PSKV also does not replace general-purpose serving systems: vLLM and SGLang remain stronger choices for production serving, while PSKV is designed for synchronous attack-optimization loops that require native autograd. 

\section{Licenses and Existing Assets}
\label{app:licenses}

We use HarmBench and WildJailbreak following their released terms and cite the corresponding papers. We evaluate publicly available model checkpoints, including Vicuna-7B-v1.5, Llama-2-7B-Chat, Llama-3-8B-Instruct, Mistral-7B-Instruct, and Qwen2.5-7B-Instruct, under their respective model licenses or terms of use. We also refer to publicly released implementations of GCG, BEAST, AmpleGCG, and AdvPrompter, and cite the original papers and repositories where applicable. Our released PSKV code will include a license file and documentation specifying permitted research use.

\section{Ethical Considerations}
\label{app:responsible-release}
This work studies an inference optimization technique for accelerating workloads that repeatedly evaluate multiple suffix candidates sharing the same prefix. In our experiments, we instantiate this workload in the context of suffix-based jailbreak evaluation, because such attacks provide a representative and practically important setting where repeated prefix processing creates substantial computational overhead. We acknowledge that jailbreak attacks and red-teaming methods can be dual-use: while they are essential for evaluating and improving the safety of large language models, they may also be misused if deployed irresponsibly.

Importantly, PSKV does not introduce any new jailbreak algorithm, vulnerability, harmful prompt construction method, or attack vector. The attack methods considered in this paper, including GCG, GCQ, AutoDAN, BEAST, AmpleGCG, and AdvPrompter, are existing methods that have already been publicly described and released by prior work. PSKV only modifies how shared-prefix inference is executed, reducing redundant computation and memory usage during repeated evaluations. Therefore, the contribution of this paper is an efficiency improvement to an existing class of workloads, rather than a new capability for generating harmful content or discovering previously unknown vulnerabilities.

We also emphasize that PSKV does not directly generate, curate, or disseminate harmful instructions, harmful completions, or new jailbreak datasets. Our method is an acceleration framework that operates at the systems level by reusing the KV cache associated with a common prefix and computing candidate-specific suffixes more efficiently. It is independent of the semantic content of the prompts being evaluated and can be applied to benign workloads with similar shared-prefix structure. In the safety evaluation setting, its primary benefit is to make red-teaming and robustness evaluation more affordable and scalable, enabling researchers and safety teams to test models more thoroughly under limited compute budgets.

At the same time, we recognize that reducing computational cost may raise concerns about misuse. To mitigate this risk, we adopt a responsible release strategy. The released code will focus on the modular PSKV acceleration framework rather than bundled end-to-end jailbreak attack pipelines. We will not include harmful prompt datasets or additional tools designed to generate harmful content. Access to the code will be governed by a research-only license that explicitly prohibits malicious use, and users will be required to acknowledge an acceptable use policy before obtaining access. This follows the precedent of responsible releases for safety evaluation resources in prior work, such as HarmBench and GCG.

Overall, the intended use of PSKV is defensive safety evaluation. By reducing the cost of large-scale jailbreak evaluation, PSKV can help model developers, auditors, and researchers perform more comprehensive safety assessments, compare defenses under broader attack settings, and identify model weaknesses before deployment. We encourage users of PSKV to apply it only in controlled research, auditing, and safety evaluation contexts, and not for generating or distributing harmful content.
\section*{NeurIPS Paper Checklist}

\begin{enumerate}

\item {\bf Claims}
    \item[] Question: Do the main claims made in the abstract and introduction accurately reflect the paper's contributions and scope?
    \item[] Answer: \answerYes{}.
    \item[] Justification: The abstract and introduction state that PSKV is an efficiency optimization for suffix jailbreak evaluation, and the scope is limited to shared-prefix suffix attacks and the evaluated open-source LLMs.

\item {\bf Limitations}
    \item[] Question: Does the paper discuss the limitations of the work performed by the authors?
    \item[] Answer: \answerYes{}.
    \item[] Justification: The paper includes a Limitations section in Appendix~\ref{app:limitation}.

\item {\bf Theory assumptions and proofs}
    \item[] Question: For each theoretical result, does the paper provide the full set of assumptions and a complete proof?
    \item[] Answer: \answerYes{}.
    \item[] Justification: The complexity assumptions and derivations are provided in Appendix~\ref{appendix:complexity} and Appendix~\ref{appendix:complexity-batch}.

\item {\bf Experimental result reproducibility}
    \item[] Question: Does the paper fully disclose all the information needed to reproduce the main experimental results of the paper to the extent that it affects the main claims and/or conclusions of the paper?
    \item[] Answer: \answerYes{}.
    \item[] Justification: The experimental setup, datasets, target models, attack hyperparameters, metrics, and compute hardware are described in Section~5 and Appendix~\ref{app:jailbreak-impl}.

\item {\bf Open access to data and code}
    \item[] Question: Does the paper provide open access to the data and code, with sufficient instructions to faithfully reproduce the main experimental results?
    \item[] Answer: \answerYes{}.
    \item[] Justification: We provide an anonymized supplementary code package with scripts, environment files, and instructions for reproducing the main experiments. The public release will follow the responsible-use policy described in Appendix~\ref{app:responsible-release}.

\item {\bf Experimental setting/details}
    \item[] Question: Does the paper specify all the training and test details necessary to understand the results?
    \item[] Answer: \answerYes{}.
    \item[] Justification: The safety dataset, model list, attack list, hyperparameters, evaluation metric, and baseline definitions are given in Section~5 and Appendix~\ref{app:jailbreak-impl}.

\item {\bf Experiment statistical significance}
    \item[] Question: Does the paper report error bars suitably and correctly defined or other appropriate information about the statistical significance of the experiments?
    \item[] Answer: \answerYes{}.
    \item[] Justification: The main ASR table reports mean $\pm$ standard deviation over three runs. Appendix~\ref{app:asr-stability} further provides paired multi-seed checks, a paired $t$-test, confidence interval, effect size, and logit-equivalence tests to assess whether PSKV changes attack effectiveness.

\item {\bf Experiments compute resources}
    \item[] Question: For each experiment, does the paper provide sufficient information on the computer resources needed to reproduce the experiments?
    \item[] Answer: \answerYes{}.
    \item[] Justification: Section~5 states that experiments were benchmarked on two 80GB A100 GPUs and reports memory and runtime costs in the main result tables.

\item {\bf Code of ethics}
    \item[] Question: Does the research conducted in the paper conform, in every respect, with the NeurIPS Code of Ethics?
    \item[] Answer: \answerYes{}.
    \item[] Justification: The work is framed as safety evaluation research, and the paper includes ethical considerations and a responsible release plan for the dual-use acceleration library.

\item {\bf Broader impacts}
    \item[] Question: Does the paper discuss both potential positive societal impacts and negative societal impacts of the work performed?
    \item[] Answer: \answerYes{}.
    \item[] Justification: The Ethical Considerations section discusses the intended safety-auditing benefits and the misuse risk from lowering the compute cost of existing jailbreak attacks.

\item {\bf Safeguards}
    \item[] Question: Does the paper describe safeguards that have been put in place for responsible release of data or models that have a high risk for misuse?
    \item[] Answer: \answerYes{}.
    \item[] Justification: Appendix~\ref{app:responsible-release} proposes a gated, research-only release without bundled harmful datasets or turnkey malicious scripts.

\item {\bf Licenses for existing assets}
    \item[] Question: Are the creators or original owners of assets used in the paper properly credited and are the license and terms of use explicitly mentioned and properly respected?
    \item[] Answer: \answerYes{}.
    \item[] Justification: The paper cites the datasets, model checkpoints, and baseline attack implementations used in the experiments. Appendix~\ref{app:licenses} summarizes the existing assets and states that their respective licenses and terms of use are followed.

\item {\bf New assets}
    \item[] Question: Are new assets introduced in the paper well documented and is the documentation provided alongside the assets?
    \item[] Answer: \answerYes{}.
    \item[] Justification: PSKV is the new code asset introduced by this work. Appendix~\ref{app:using-pskv} documents the integration interface and usage pattern, and Appendix~\ref{app:responsible-release} describes the responsible release policy and required usage restrictions.

\item {\bf Crowdsourcing and research with human subjects}
    \item[] Question: For crowdsourcing experiments and research with human subjects, does the paper include the full text of instructions given to participants and screenshots, if applicable, as well as details about compensation?
    \item[] Answer: \answerNA{}.
    \item[] Justification: The work does not involve crowdsourcing or new human-subject experiments.

\item {\bf Institutional review board approvals or equivalent for research with human subjects}
    \item[] Question: Does the paper describe potential risks incurred by study participants, whether such risks were disclosed to the subjects, and whether IRB approvals or equivalent were obtained?
    \item[] Answer: \answerNA{}.
    \item[] Justification: The work does not involve new human-subject experiments.

\item {\bf Declaration of LLM usage}
    \item[] Question: Does the paper describe the usage of LLMs if it is an important, original, or non-standard component of the core methods in this research?
    \item[] Answer: \answerYes{}.
    \item[] Justification: LLMs are the target models and, for model-based attacks, part of the evaluated attack pipelines; these uses are described in Section~5 and Appendix~\ref{app:jailbreak-impl}. We do not use LLMs to generate new benchmark labels or human-subject data.

\end{enumerate}
\end{document}